\documentclass[aps,prx,reprint,groupedaddress,longbibliography]{revtex4-1}

\usepackage{amsmath}
\usepackage{graphicx}
\usepackage{color}
\usepackage{bbold}

\begin{document}

% Use the \preprint command to place your local institutional report
% number in the upper righthand corner of the title page in preprint mode.
% Multiple \preprint commands are allowed.
% Use the 'preprintnumbers' class option to override journal defaults
% to display numbers if necessary
%\preprint{}

\title{    Approach  to equilibrium and non-equilibrium stationary distributions 
of  interacting many-particle systems that are coupled to different heat baths
  }

% \title{ \textcolor{red}{   Approach of interacting   many-particle systems  that are coupled to different heat baths
 %to equilibrium and non-equilibrium stationary distributions   }}

%\title{ Derivation of the  functional that governs  the  approach to equilibrium and non-equilibrium stationary distributions  for
%interacting   many-particle systems }

% repeat the \author .. \affiliation  etc. as needed
% \email, \thanks, \homepage, \altaffiliation all apply to the current
% author. Explanatory text should go in the []'s, actual e-mail
% address or url should go in the {}'s for \email and \homepage.
% Please use the appropriate macro foreach each type of information

% \affiliation command applies to all authors since the last
% \affiliation command. The \affiliation command should follow the
% other information
% \affiliation can be followed by \email, \homepage, \thanks as well.
\author{Roland R. Netz}
\email[]{rnetz@physik.fu-berlin.de}
%\homepage[]{Your web page}
%\thanks{}
%\altaffiliation{}
\affiliation{Fachbereich Physik, Freie Universit\"at Berlin, 14195 Berlin, Germany}

%Collaboration name if desired (requires use of superscriptaddress
%option in \documentclass). \noaffiliation is required (may also be
%used with the \author command).
%\collaboration can be followed by \email, \homepage, \thanks as well.
%\collaboration{}
%\noaffiliation

\date{\today}

\begin{abstract} % less than 500 words
A Hamiltonian-based   model of  many  harmonically interacting  massive particles that 
  are subject to   linear friction 
 and coupled to heat baths at different temperatures  is used to 
study  the dynamic approach to  equilibrium and non-equilibrium stationary states.
%Interactions, friction couplings  and stochastic fields  are in the  first part assumed completely general.
%
An  equilibrium system is  here defined as a system whose 
stationary distribution   equals  the  Boltzmann distribution,
 the relation of this definition to the conditions of  detailed balance
 and vanishing probability current  is discussed  both for underdamped as well as for overdamped systems.
Based on the exactly calculated  
 dynamic approach   to the stationary distribution,
the functional that governs this  approach, which is called  the free entropy
${\cal S}_{\rm free}(t)$, is constructed.
For the stationary distribution ${\cal S}_{\rm free}(t)$   becomes
 maximal and  its time derivative, the free entropy 
production $\dot{\cal S}_{\rm free}(t)$, is minimal and vanishes. 
%The general extremal and stability conditions we  derive 
%for the free entropy  describe equilibrium as well as non-equilibrium systems. 
Thus, ${\cal S}_{\rm free}(t)$  characterizes equilibrium as well as non-equilibrium  stationary 
distributions by their  extremal and stability properties.
For an equilibrium system,   i.e. if all heat baths have the same temperature,
  the free entropy equals the negative  free energy divided by temperature
  and thus corresponds to the Massieu function which was previously introduced  in an alternative
  formulation of statistical mechanics.
%As a byproduct of our approach, 
%we derive the necessary and sufficient condition for the many-body Langevin
%equations to describe an equilibrium system.
Using  a systematic perturbative scheme for calculating 
velocity and position correlations
in the overdamped massless limit,   explicit  results  for  few particles are presented:
 For two particles  localization in position and momentum  space is demonstrated
 in the  non-equilibrium stationary state, indicative of a tendency to phase separate.
 For three elastically interacting particles 
    heat  flows from a particle coupled to a  cold reservoir to a particle coupled to a warm reservoir  if
 the third reservoir is sufficiently hot. 
This does not constitute a violation of the second law of thermodynamics, but rather demonstrates that a particle
 in such a  non-equilibrium system  is not  characterized  by an effective temperature  which equals  the temperature of the heat bath  it is coupled to.
Active particle models  can be described in the same general framework,
which  thereby allows to characterize their entropy production not only in  the stationary state
but also in the approach  to the stationary non-equilibrium state. 
 Finally,  the  connection to non-equilibrium thermodynamics formulations that 
include the reservoir entropy production is discussed. 

\end{abstract}

% insert suggested PACS numbers in braces on next line
% insert suggested keywords - APS authors don't need to do this
%\keywords{}

%\maketitle must follow title, authors, abstract, \pacs, and \keywords
\maketitle

% body of paper here - Use proper section commands
% References should be done using the \cite, \ref, and \label commands
%\section{}
% Put \label in argument of \section for cross-referencing
%\section{\label{}}
%\subsection{}
%\subsubsection{}

%\tableofcontents

\section{Introduction}

Systems that even in their stationary state are  not in equilibrium 
have  in the last decades received renewed  attention since standard concepts of 
thermodynamics and statistical mechanics do not work and because of their  experimental relevance \cite{Onsager1931,Prigogine1947,Mazur1953,Lebowitz1959,Graham1971,Schloegl1971,Procaccia1976,Schnakenberg1976,Oono1998,Hatano2001,Esposito2007,RamaswamyReview,ZiaReview,SeifertReview}.
Some of the motivation for studying such systems comes from 
experimental observations of molecular
 processes in living  systems, which are 
fundamentally  non-equilibrium (NEQ)\cite{Marchetti2013,BroederszReview}.
Apart from biological applications, 
experimental advances allow for the construction of  NEQ systems 
either from biological components that are driven out of equilibrium
by ATP consumption or by using  macromolecular and colloidal assemblies 
that can be driven out of equilibrium by  chemical fuels or by applying external 
forces\cite{LoewenReview,Golestanian2017}.

Two  prominent features of  NEQ systems
are departures from the canonical Boltzmann distribution 
and   violation of the fluctuation-dissipation theorem, 
which were recently demonstrated to go hand in hand for a harmonically coupled particle system \cite{Netz2018},
 the fundamental  hallmark of NEQ systems is the violation of the detailed balance condition \cite{Graham1971,Schnakenberg1976}.
A particularly striking example for the breakdown of equilibrium statistical mechanics 
is the occurrence of phase transitions induced by  NEQ driving. 
As a result of external forces acting on particle systems,  laning and 
 phase separations have been observed
\cite{Spohn,Krug,Zia,Dzubiella,Netz2003,Ramaswamy}.
Internal  NEQ effects can generate  particle propulsion that is sustained by
 chemical or biochemical means. NEQ effects can also be induced
by  stochastic forces due to  the coupling of particles 
to different heat or chemical energy reservoirs. Such internal NEQ effects
 have also been shown to lead to clustering
and phase separation \cite{Bertin,Bocquet2012,Golestanian2012,Loewen2014,Bocquet2015,Grosberg2015,Frey,Brenner2017,Kremer2017,Prabal2019}.
Experimentally,  symmetry-breaking transitions in suspensions of swimming bacteria and
 filament  systems driven by motor proteins
 have indeed been demonstrated \cite{Sokolov,Bausch}.
Also in  more coarse-grained models, defined by  migration rules,
collective ordering can be  obtained and describes the
response of pedestrians to spatial confinement
or the flocking and swarming of animals \cite{Helbing,RamaswamyReview}.
NEQ   transformations also occur for  polymers   that
are driven by externally applied torques \cite{Wolgemuth,Netz2009}.

The  equilibrium fluctuation-dissipation theorem (FDT) 
describes  the dynamical response of an equilibrium  system
to a small perturbation. Since it is  a key concept for systems close to equilibrium,
and since any violation of the FDT clearly signals that a system is off equilibrium, 
the derivation of generalized fluctuation-dissipation relations that also hold for  
NEQ systems is  central and has been 
 discussed in the context of  laser \cite{Agarwal1972}, chaotic  \cite{Shraiman1989},
 glassy  \cite{Kurchan1997},  driven colloidal \cite{Seifert2006}, 
 sheared  \cite{Hern2005,Barrat2007,Fuchs2009}
 and active systems \cite{Levis2015}.
 Generalized NEQ  fluctuation-dissipation relations were derived 
 \cite{Agarwal1972,Harada2005,Zamponi2005,Prost2009,Wynants2009,Seifert2010,Lindner2017} 
and compared with experimental data  for glasses  \cite{Bonn08},  
colloids \cite{Gawedzki2009,Seifert2010b}, bundles of  biological filaments
 \cite{Julicher2001,Joanny2012}, living cells \cite{Bohec2013}
  and biological gels that are driven by motor proteins 
 in the presence of ATP\cite{Schmidt07,Netz2018}.

 Significant  theoretical progress has been made in the characterization
of NEQ systems by generalized NEQ fluctuation-dissipation relations, as discussed above, 
and by  fluctuation relations 
 that give bounds and exact relations involving  trajectory ensembles \cite{Lebowitz1999,Crooks1999,Jarzynski2000}.
  The field of stochastic thermodynamics has linked the statistics of
trajectories to  entropy-production contributions \cite{Seifert2005,SeifertReview}.
A good fraction of contemporary  work on NEQ systems is 
concerned with  analyzing the solutions of  
governing dynamic equations either by simulation techniques  or, for simple systems, 
by analytical methods.
In essence, and as acknowledged by most  workers in the field,  NEQ theory is
far from the predictive power and  understanding furnished  by  e.g. the
usage  of  thermodynamic potentials in the context of  equilibrium scenarios.
 For an isolated system the entropy is maximized,
 %which holds for equilibrium as well NEQ systems,  
 but this provides little help for the type of NEQ  problems one is
 typically interested in, because they are not isolated. 
 For a system 
coupled to a single heat reservoir and in the absence of external driving forces, which gives rise to  the equilibrium canonical ensemble,
 the free energy  is minimized. 
The free energy, when evaluated exactly or approximately, 
allows to predict phase transitions, structures and all static properties
of an equilibrium  system. In the search for a similarly useful framework, the first  theoretical studies on NEQ systems formulated  general extremal principles 
that  express  the system's tendency to extremize its dissipation, i.e., its entropy 
 production \cite{Prigogine1947,Mazur1953}.
 These early  extremal principles were  limited to linear 
 and homogeneous systems close to equilibrium and included  interactions
 only indirectly  in terms of  phenomenological coefficients.
  Subsequently, the Master equation approach allowed to derive 
  extremal principles for general  NEQ stationary states in terms of a generalized entropy \cite{Graham1971,Schloegl1971,Schnakenberg1976, Esposito2007}.
 More recently, Onsager's  variational principle\cite{Onsager1931}
 was revisited and used for the study of various NEQ problems in soft condensed matter \cite{Doi2019}.
 
 In this paper we  start from  a
quadratic Hamiltonian model  with general interactions. 
By adding  friction terms and stochastic fields to the Hamilton equations,
we arrive at the  general linear Hamiltonian-based  many-dimensional Langevin equation
 that, for suitably chosen friction and stochastic parameters,
  describes a  many-body system of massive particles  coupled to multiple heat baths with different and well-defined
   temperatures,
a model that corresponds to the non-equilibrium version of the multidimensional Ornstein-Uhlenbeck process
and has in certain limits been studied  in literature
 \cite{Lebowitz1959,Lebowitz1967,Zwanzig1973,Filliger2007,Prost2009,Ramaswamy2018,Li2019}.
  The explicit presence of  heat baths with different  temperatures allows 
  us to prepare the system in  a unique NEQ stationary state and to calculate all contributions to the entropy production as the system approaches
  the stationary state.
 Instead of considering trajectories in phase space, we base our theory on the time-dependent distribution. 
 The key point  of our model  is that we can exactly calculate the time derivative of the
  distribution entropy 
and from that  construct, by comparison with the independently calculated relaxation of the distribution, 
 the time-dependent  functional that governs the  approach to equilibrium as
well to NEQ stationary distributions. 
In analogy to the relation between  the  energy and the free energy,
this functional is called  the free entropy,
${\cal S}_{\rm free}(t)$,
since it contains the distribution entropy of the system 
%but not the  reservoir entropy 
and also  accounts for  the interactions within the system and the coupling to the reservoirs.
Using the free entropy ${\cal S}_{\rm free}(t)$,
 %the time derivative of 
 the  total entropy ${\cal S}_{\rm tot}(t)$, wich includes the entropy of  the
interacting particle system $ {\cal S}(t)$ as well as the reservoir entropy ${\cal S}_{\rm res}(t)$, can be decomposed as
\begin{equation} 
\label{intro1}
%\dot{\cal S}_{\rm tot}(t) =   \dot{\cal S}_{\rm free} (t)+ \dot{\cal S}^\circ_{\rm res},
{\cal S}_{\rm tot}(t) =  {\cal S}(t)+{\cal S}_{\rm res}(t)= {\cal S}_{\rm free} (t)+ t \dot{\cal S}^\circ_{\rm res}
  \end{equation}
up to unimportant constants,
where $ \dot{\cal S}^\circ_{\rm res}$ denotes the  entropy production due to heat transfer
with all reservoirs in the unique stationary state.
While ${\cal S}_{\rm res}(t)$ is difficult to calculate for the stochastic models used for the description 
of heat reservoirs and in fact increases boundlessly, $ {\cal S}_{\rm free} (t)$ and $\dot{\cal S}^\circ_{\rm res}$ can be explicitly calculated.
 Similar to the free energy of statistical mechanics,  different  observables 
 can be derived from the free entropy by taking suitable derivatives, as is shown in 
 Sect.  \ref{extremalprop}.
 
For an  equilibrium system,
i.e. when all heat baths have the same temperature and $ \dot{\cal S}^\circ_{\rm res}=0$,
and in the stationary state,
 the free entropy is time-independent and equals the negative  free energy divided by temperature, i.e.
 \begin{equation}
\label{Sfree0}
 {\cal S}^{\bullet}_{\rm free}   = -  { \cal F}^{\bullet}/T   =   {\cal S}^{\bullet} -  { \cal U}^{\bullet}/T,
\end{equation}
where  filled circles in our paper denote the equilibrium stationary state.
 In fact, the free entropy  functional had been used by Massieu already in 1869 \cite{Callen,Balian},
 a few years before Gibbs  introduced his energy transforms.
 The advantage of the free entropy  functional has been pointed out by Planck \cite{Planck} and 
 Schrödinger \cite{Schroedinger}, while the name was introduced more recently in the mathematical literature \cite{Biane2001,Voiculescu2002}.
 The free entropy is central in the context of  our model, since the presence of different heat bath temperatures 
 does not allow to define a unique NEQ version of the free energy. 
%
%The stationary  distribution is given by the maximum 
%of the free entropy, and stability follows from the fact that the free entropy
 %production is positive
%except for the stationary  distribution, for which the free entropy production vanishes. 
%For the stationary distribution, the total entropy production is thus given solely by 
%the reservoirs. 
%
%The concept  of our paper is similar to one previous work,
%where from the exact solution of the diffusive transport equations between two reservoirs,
%a NEQ free energy functional has been derived\cite{Derrida2001}.
In a number of previous papers functionals were derived that in a NEQ stationary state 
are extremal and  thereby allow to study the  relaxation and the stability of NEQ systems
and the relation of these functionals  to the Kullback-Leibler entropy\cite{Kullback1951} was pointed out
\cite{Graham1971,Schloegl1971,Procaccia1976,Schnakenberg1976,Kwon2005,Esposito2007,Ge2009,Ge2010,Qian2013,Qian2015}.
 Our model differs from those works since we introduce NEQ by the coupling to multiple heat baths with different temperatures.
While the early approaches to NEQ thermodynamics were centered on the total entropy production $\dot{\cal S}_{\rm tot}(t)$,
i.e. the time derivative of the total entropy including the reservoirs \cite{Prigogine1947,Mazur1953},
  in later developments the entropy production due to heat transfer from the reservoirs,
which   in a NEQ stationary state is constant and given by $ \dot{\cal S}^\circ_{\rm res}$,
has been separated off and  called the house-keeping entropy  \cite{Oono1998,Hatano2001,Speck2005}.
 One advantage of our model is that 
 since the particles have finite masses,
  the heat fluxes between the particles and the heat reservoirs can be calculated from 
  energy balance considerations including the kinetic energy and thus the stationary reservoir entropy production 
  $ \dot{\cal S}^\circ_{\rm res}$ can be derived   straightforwardly.
  For this we introduce a systematic    perturbation scheme
  using the particle masses as expansion parameters
       to calculate NEQ mixed position-velocity  correlations.

We here show that for a system coupled to different temperature reservoirs, 
the free entropy functional ${\cal S}_{\rm free}(t)$  can be written down explicitly  and is for  the NEQ stationary distribution
 maximal and constant  in time. 
The free entropy production  $\dot{\cal S}_{\rm free}(t)$ 
is positive except at the stationary NEQ state,
for which it  vanishes, i.e. 
\begin{equation}
\dot{\cal S}_{\rm free}(t) \geq 0.
\end{equation}
This  shows
that the NEQ stationary state is stable with respect to small perturbations and
that in the NEQ stationary state
 the total entropy production is  given solely by the reservoirs. 
The time derivative of the free entropy  production,
which would be the second time derivative of the free entropy,
  is not needed, unlike early NEQ approaches \cite{Prigogine1947,Mazur1953}.
Our framework thus treats NEQ and equilibrium systems 
on the same footing, as for an equilibrium system  the NEQ  free entropy   
 smoothly crosses over to the standard  free energy divided by $-T$,
 see Eq. (\ref{Sfree0}),
 and so the  standard equilibrium and stability conditions of  
 statistical mechanics are recovered.

Interestingly, the dynamic approach to  the equilibrium 
and  to the NEQ stationary distributions obey the same differential equation, 
 as is shown in Sect. \ref{DynApproach}.
In fact, while equilibrium and NEQ stationary distributions exhibit many fundamental differences,
the relaxation times that characterize the approach  to stationarity  
are independent of the heat bath temperatures and  therefore
do not allow to distinguish equilibrium  from  NEQ systems.
While this might be a simplification due to our neglect of non-linear
interactions, we argue  that  non-linear systems can typically  be quadratically approximated around
locally stable states  and thus our results  should also apply  to sufficiently well-behaved 
non-linear systems.

We also demonstrate that the definition of equilibrium we are using in this paper,
which is based on the Boltzmann distribution,  is 
equivalent to the detailed-balance condition only 
if  the friction matrix is  symmetric and if  the random fields couple separately to position and velocity 
 degrees of freedom.
For a simple system consisting of  two coupled massive particles, we show that if the 
Boltzmann distribution is realized, the fluctuation-dissipation theorem is satisfied, even when 
the friction matrix is asymmetric (and thus   the condition of detailed balance is not satisfied). 
 This shows that equilibrium definitions based on Boltzmann statistics, on the fluctuation-dissipation theorem
 and on the detailed-balance condition 
are   equivalent only  for symmetric friction matrices and that the detailed-balance condition is the strictest of all three.

As a simple application of our model we present results for three interacting massive particles
that are coupled to temperature reservoirs at different temperatures,
for which  we demonstrate that heat  flows
 from a particle coupled to a cold reservoir to a particle coupled to a  warm reservoir
if the third reservoir is sufficiently hot. This  of course does not constitute 
a violation of the second law of thermodynamics. Rather, this NEQ
 entrainment effect can be rationalized by the fact that
 the reservoirs are not coupled directly to each other but rather indirectly via the particles,
 and that the particles  are not characterized solely by  the heat bath temperatures.
 This point can be explained in more detail by considering just two 
 particles that are coupled to different heat baths:
 We demonstrate that  the concept of a 
 NEQ effective  temperature has only a rather limited value,
  since each covariance matrix entry of two coupled NEQ  particles would have
 to be attributed a different effective temperature. 
 For two particles we also demonstrate that NEQ effects 
 give rise to localization effects both in position and in momentum space,
 which is reminiscent of  attractive interactions.  This reflects  the tendency of 
 NEQ systems to phase separate in both position and momentum space. 
 Finally, we show how active particle models can be described using our general framework
 and discuss the connection between  the free entropy production and the total entropy production
 that includes the reservoirs. 
 
 In the following sections we first treat general NEQ systems and derive the necessary conditions to 
 reach a stationary state and a stationary equilibrium state, described by the Lyapunov and
 the Lyapunov-Boltzmann equations, respectively. 
 In  Sect. \ref{EntropyProd} we start treating the core model of this paper, where  particles are coupled 
 to heat reservoirs that are characterized by different temperatures, and present various explicit examples.

\section{Many-Particle Hamiltonian Model}

\subsection{From Hamilton to Langevin equations}

To proceed,  we consider 
$N$ massive particles in one dimension  with positions $x_\alpha$ and momenta $p_\alpha$
that move according to the Hamilton equations
\begin{equation}
\dot{x}_\alpha(t) = \frac{\partial {\cal H} ( \vec{x}, \vec{p})}{\partial p_\alpha},
\end{equation}
\begin{equation}
\dot{p}_\alpha(t) = - \frac{\partial {\cal H}  ( \vec{x}, \vec{p}) }{\partial x_\alpha},
\end{equation}
where $\alpha = 1 \ldots N$ is an index that   runs over all particles.
The case of   $M$ interacting particles
in three dimensions is described by $N=3M$ particle coordinates
and is implicitly included in our model. 
Using the antisymmetric matrix 
 \begin{equation} \label{U}
U = \begin{pmatrix}
 0 & 1  &0  &0 & \\
-1  & 0 & 0 & 0 & \\
 0 & 0 &0  &1 & \\
0 & 0 & -1 & 0 & \\
    &    &    &    & \ddots 
\end{pmatrix}
\end{equation}
and the state vector 
 \begin{equation}
\vec{z}(t) = (x_1(t), p_1(t), x_2(t), p_2(t) \ldots)^T
\end{equation}
the Hamilton equations can be written compactly as
 \begin{equation}
\dot{z}_i(t) = U_{ij}\frac{\partial {\cal H} (\vec{z})  }{\partial z_j},
\end{equation}
where $i = 1 \ldots 2N$ is an index that   runs over all position and momentum coordinates.
Throughout this paper, greek indices denote particles (running from 1 to $N$),
roman indices denote coordinates (running from 1 to $2N$)
and indices that appear more than once are summed over except
primed indices.
We consider  quadratic Hamiltonians of the form 
 \begin{equation}
 {\cal H} (\vec{z})   =  z_i H_{ij}  z_j /2
\end{equation}
that are described  by a general symmetric matrix $H$ which we assume to be
positive definite, i.e., $ {\cal H} (\vec{z}) > 0$ for general $\vec{z}$. 
A possible linear term in the Hamiltonian can be absorbed into the definition of the state vector 
$z_j$ and need  not be considered explicitly.
For quadratic Hamiltonians  the Hamilton equations are linear and   given by
 \begin{equation}
\dot{z}_i(t) = U_{ij}  H_{jk}    z_k(t).
\end{equation}
More specific forms of the Hamiltonian matrix $H$, 
in particular Newtonian Hamiltonians  where momentum und position degrees 
of freedom are decoupled,  will be discussed later, 
in the first part of this paper the discussion applies to general Hamiltonian models.

By adding linear  friction  terms and random fields  to the Hamilton equations, 
which will be later shown to mimic the coupling  to heat   baths with in general different temperatures,
we obtain  the coupled linear Langevin equations 
\begin{equation}
	\dot{z}_i(t)  =  U_{ij}  H_{jk}     z_k(t)  -\Gamma_{ik} z_k(t) + \Phi_{ik} F_k  (t)
\end{equation}
which by definition of the  generally asymmetric coupling matrix
\begin{equation}
	\label{A}
	A_{ik}   =  -  U_{ij}  H_{jk}   + \Gamma_{ik}
\end{equation}
can be written more compactly as
\begin{equation}
	\label{Langevin}
\dot{z}_i(t)= -A_{ik} z_k(t) + \Phi_{ik} F_k  (t).
\end{equation}
Here,  $\Gamma$ is the friction coefficient matrix and  $\Phi$ is the random  strength matrix that describes how random 
fields  couple to different particle coordinates. 
For simplicity, we assume Gaussian white random fields with zero mean $\langle F_i  (t) \rangle = 0$ and   and variances 
$\langle F_i  (t) F_j (t')\rangle = 2\delta_{ij} \delta(t-t')$.
Non-Markovian models with colored noise can be obtained by integrating out degrees of freedom and 
need not   explicitly be considered \cite{zwanzig_memory_1961}.
In  standard friction models  friction forces couple to the momentum degree of freedom 
and  are proportional to  particle velocities.
In more elaborate models that include hydrodynamic interactions,  the  friction force acting on a given particle 
depends on the  velocities of all  particles.
In the first part of this paper   the matrices $\Gamma$ and $\Phi$ are kept general 
 and $\Gamma$  can also be asymmetric (which will be shown to have direct consequences for the detailed-balance condition
 in Sect. \ref{detbal}). 
In the second part of the paper, starting in   Sect. \ref{EntropyProd},
we model heat baths with different temperatures and  for this will assume $\Gamma$ and $\Phi$
to be diagonal.
The general form of the linear Langevin Eq. (\ref{Langevin}) does not directly  reveal whether it 
describes an equilibrium or a NEQ  system\cite{Ramaswamy2018}, 
this point will be addressed further below.

\subsection{Stationary distribution} \label{stationary}

The algebraic  solution of the Langevin Eq. (\ref{Langevin})  is 
 \begin{equation}
	\label{sol1}
z_i (t)  = {\rm e}^{ - t A }_{ij} z_j(0) +  \int_0^t {\rm d} t'   {\rm e}^{ - (t-t') A }_{ij}  \Phi_{jk} F_k  (t'),
\end{equation}
where $z_j(0)$ denotes the initial particle positions and momenta at time zero.
The average over the noise gives
 \begin{equation}
	\label{sol2}
\langle z_i (t) \rangle   = {\rm e}^{ - t A }_{ij} z_j(0) 
\end{equation}
which can be viewed as the  solution of the noise-averaged version of the Langevin Eq. (\ref{Langevin}) 
\begin{equation}
\label{avLangevin}
\langle \dot{z}_i(t) \rangle = -A_{ik} \langle z_k(t) \rangle.
\end{equation}
If all eigenvalues of   the matrix $A$ have positive real  components,
a unique   stationary distribution  exists and is characterized  by a vanishing mean
$\langle z_i (t \rightarrow \infty ) \rangle =0$.

The covariance matrix of the deviations from the mean
$\Delta z_i (t) = z_i (t)  - \langle z_i (t) \rangle$ follows from 
squaring the solution Eq. (\ref{sol1}) and averaging over the noise, 
leading to\cite{Risken}
\begin{equation}
	\label{sol3}
E_{ij} (t)  \equiv   \langle  \Delta z_i (t)  \Delta z_j (t)    \rangle
= 2   \int_0^t {\rm d} t'  {\rm e}^{ - (t-t') A }_{ik}    {\rm e}^{ - (t-t') A }_{jl }    C_{kl},
\end{equation}
where the random correlation matrix is defined by 
\begin{equation} \label{CCC}
C_{kl}=\Phi_{km}\Phi_{lm}=C_{lk}
\end{equation}
and  is symmetric by construction.
Since 
\begin{eqnarray}
\label{Lyapderive}
A_{ik}E_{kj}+A_{jk}E_{ki}=A_{ik} E_{kj}+   A_{jk} E_{ik}  =  \nonumber \\
2\int_0^t{\rm d}t'\frac{\rm d}{{\rm d}t'}{\rm e}^{-(t-t')A}_{ik}{\rm e}^{-(t-t')A }_{jl}C_{kl},
\end{eqnarray}
the stationary covariance matrix, denoted by an open circle and defined by 
\begin{equation}
	\label{stat}
E^{\circ}_{ij}  \equiv   E_{ij} (t \rightarrow \infty ),
\end{equation}
 is unique and given by the Lyapunov equation 
\begin{equation}
\label{Lyapunov}
2C_{ij}=A_{ik} E_{kj}^{\rm \circ} +   A_{jk} E_{ki}^{\rm  \circ}
\end{equation}
 if all eigenvalues of the matrix $A$ have positive real  parts,
 which we will assume to be true throughout this paper.

%%%%%%%%%%%%%%%%%%%%%%%%%%%%%%%%%%%%%%%%%%%%%%%%%
\section{Systems that have an equilibrium distribution}

\subsection{Distributions, entropy and  free energy }

In  equilibrium,  the normalized  distribution  in terms of the state vector $ \vec{z}$
is given by the Boltzmann distribution 
\begin{equation} \label{Boltz}
\rho(\vec{z}) = {\rm e}^{ -  \beta {\cal H} (\vec{z})} / {\cal Z},
\end{equation}
where $\beta = 1/(k_BT)$ denotes the inverse thermal energy and 
${\cal Z}= \int {\rm d} \vec{z}{\rm e}^{ -  \beta {\cal H} (\vec{z})}$ is the partition function.
We will discuss  the connection of this definition of equilibrium to 
 the conditions of detailed balance,
  vanishing probability current as well as  the fluctuation-dissipation theorem in Sect. \ref{detbal}.
Positive definiteness of the Hamiltonian matrix $H$ guarantees that ${\cal Z} $ is finite
(if the Hamiltonian is invariant with respect to one or few degrees of freedom they can be 
 separated off to make the reduced Hamiltonian positive definite).
 For a quadratic Hamiltonian, 
 the  average state vector  vanishes
 and all covariances can be calculated  from the Boltzmann
 distribution, which only involves  inversion of the Hamiltonian matrix. 
 
 For the later discussion  of the NEQ scenario, it is instructive to derive the equilibrium 
 distribution also via the thermodynamic route. 
 From the thermodynamic definitions of the free energy ${\cal F}  =- k_BT  \ln {\cal Z} $ and the entropy
${\cal S}  = - \partial {\cal F}  /\partial T$, the Shannon expression for the entropy directly follows as
\begin{equation}
\label{entropy}
{\cal S}  /k_B   = -   \int {\rm d} \vec{z}  \rho(\vec{z})  \ln \rho(\vec{z}),
\end{equation}
the derivation is shown in Appendix \ref{AppA}. 
Note that the Shannon expression Eq. (\ref{entropy}) can also be used  to describe the distribution entropy   for 
time-dependent distributions,
 i.e. for non-stationary and even NEQ  situations, since the expression
 makes no reference to the equilibrium ensemble or to the presence of a heat bath.  
This  will allow  us to describe the time-dependent approach to equilibrium as well
as  to stationary NEQ distributions.

For the linear  Langevin Eq. (\ref{Langevin}),  the time-dependent probability distribution is  Gaussian and 
can be written as 
\begin{eqnarray}
	\label{rho}
\rho( \vec{z},  %\langle  \vec{z} (t) \rangle,     
 t )  =    \hspace{5cm}\\
{\cal N}^{-1} (t)  \exp\left(-    [z_i  - \langle z_i (t) \rangle ]      
 E^{-1}_{ij}(t)  [z_j  - \langle z_j (t) \rangle]   /2\right),
\nonumber  \end{eqnarray}
where the time-dependent normalization constant is given by 
\begin{equation}
	\label{N}
 {\cal N}(t)  = \sqrt{(2\pi)^{2N} \det E(t)}
\end{equation}
and only exists if the covariance matrix  $E(t)$  is positive definite 
(note that by its definition  $E(t)$ is also symmetric).
   The proof is standard textbook material \cite{Risken}, 
in  Appendix  \ref{AppB} we present a derivation based on random-field
path integrals, which has the advantage that it can in principle be generalized to 
non-Gaussian colored noise; there we  furthermore demonstrate that the expression Eq. (\ref{rho})
is in fact the Green's function of the general Langevin Eq. (\ref{Langevin}), i.e., the conditional probability distribution at time $t$ for the case
that the distribution is a delta function at time $t=0$. 

With  the Gaussian  form Eq. (\ref{rho}), the integral in Eq. (\ref{entropy}) can be performed
and yields the
time-dependent distribution entropy as 
\begin{equation}
{\cal S} (t) /k_B   =  N + \ln  {\cal N}(t) = N +  (1/2)\ln[(2\pi)^{2N}\det E(t) ].
\end{equation}
The internal energy is given by 
\begin{eqnarray}  \label{internalenergy}
{\cal U} (t)  & =   \langle {\cal H}(\vec{z}(t)) \rangle = H_{ij}  \langle z_i(t)  z_j(t)  \rangle /2 \\
&= H_{ij}   \langle z_i(t)  \rangle  \langle   z_j(t)  \rangle /2  + H_{ij} E_{ij}(t)  /2. \nonumber
\end{eqnarray}
With these results for the entropy and internal energy, the free energy  
\begin{equation}
\label{freeen}
{\cal F} (t)   =   {\cal U} (t) -  T {\cal S} (t) 
\end{equation}
follows as 
\begin{eqnarray} \label{freeenb}
{\cal F} (t)   = & H_{ij}   \langle z_i(t)  \rangle  \langle   z_j(t)  \rangle /2  + H_{ij} E_{ij}(t)  /2 \\
& -k_BT N -  (k_BT /2)\ln [ (2\pi)^{2N}\det E(t)]. \nonumber
\end{eqnarray}
 The extremum of  the free energy is  determined  by  $\langle \vec{z}(t)   \rangle=0$  and by 
\begin{equation}
	\label{extremize}
\frac{  \partial {\cal F} (t) }{\partial E_{ij}}   =  
H_{ij}/2 - k_BT E^{-1}_{ji}(t) /2=0,
\end{equation}
the solution of which is time-independent and defines the equilibrium distribution  
(denoted by a filled circle) as 
\begin{equation}
\label{Ebullet}
 E_{ij}^{\bullet}=  k_BT H^{-1}_{ij}.
\end{equation}
In deriving Eq. (\ref{extremize}) we used the basic algebraic relation 
$ \partial  \ln \det E /  \partial E_{ij} = E^{-1}_{ji}$.
 The partial derivative  denotes the derivative with respect to one matrix component while keeping
 all other components fixed.
The equilibrium free energy follows by reinserting $\langle \vec{z}(t)   \rangle=0$
and the solution $E_{ij}^{\bullet}$ into the free energy expression Eq. (\ref{freeenb})  and  is given by 
\begin{equation} \label{Fbullet}
{\cal F}^{\bullet}=   -  (k_BT /2)\ln [ (2\pi k_BT )^{2N}/\det H].
\end{equation}

We next want to show that the extremum of the free energy is in fact a minimum
(for this we neglect the trivial quadratic dependence 
of Eq. (\ref{freeenb})  on the  mean state vector $\langle \vec{z}(t)   \rangle$).
We first realize that 
\begin{equation}
\frac{  \partial^2 {\cal F} (t) }{   \partial E_{kl}   \partial E_{ij}  }   =  
 k_BT E^{-1}_{ik}(t) E^{-1}_{lj}(t)      /2,
\end{equation}
where we used the basic algebraic relation 
$ \partial  E^{-1}_{kn} /  \partial E_{ij} = - E^{-1}_{ki} E^{-1}_{jn}$.
Around the equilibrium distribution   $E_{ij} = E^{\bullet}_{ij}$ 
 the free energy is to second order  given by
 \begin{eqnarray} \label{freeen2b}
& {\cal F} (t) - {\cal F}^\bullet   \simeq    \\
&H_{ki}   (E_{ij}(t)-H^{-1}_{ij}k_BT)  H_{jl} (E_{lk}(t)- H^{-1}_{lk} k_BT) /(2 k_BT ), \nonumber
\end{eqnarray}
which can be rewritten as 
 \begin{equation} \label{freeen2}
 {\cal F} (t) -  {\cal F}^\bullet  \simeq  
\frac{ k_BT  (\delta_{il}-\beta H_{ik}E_{kl}(t))(\delta_{il}-\beta H_{lk}E_{ki})}{2}.
\end{equation}
The latter form  is quadratic and of the general form
 ${\cal F} (t)  -   {\cal F}^\bullet \simeq  k_B T B_{il}B_{li}/2$
 with $B_{il} \equiv \delta_{il}-\beta H_{ik} E_{kl}(t)$,
 but this  by itself does  not guarantee 
that ${\cal F} (t)  -   {\cal F}^\bullet$ is positive
since $B_{il}$ is not necessarily symmetric.
In Appendix \ref{AppC} we show by diagonalization  that the positivity 
of the expression Eq. (\ref{freeen2b}) for
${\cal F} (t)  -   {\cal F}^\bullet$ follows 
 from the fact that $H$ is  symmetric and positive definite.
 
 The Gaussian distribution Eq. (\ref{rho}) in conjunction with Eq. (\ref{Ebullet})
 is equivalent to the Boltzmann distribution Eq. (\ref{Boltz}), which we 
 have thus rederived by minimizing the time-dependent free-energy functional Eq. (\ref{freeenb}).
 But the free energy functional Eq. (\ref{freeenb}) is not only valid in  equilibrium
 but also describes  systems that approach the equilibrium distribution.
 This is an  important insight,  as this functional framework will  allow us to 
 characterize the approach not only to equilibrium but also to stationary NEQ distributions.

%%%%%%%%%%%%%%%%%%%%%%%%%%%%%%%%%%%%%%%%%%%%%%%%%
\subsection{When does a  Langevin equation describe an equilibrium system?}

In this section we will explore under which conditions the Langevin Eq. (\ref{Langevin})
 describes an equilibrium system,
which will put stringent conditions on the random  correlation matrix $C$ and on the friction matrix $\Gamma$. 
We in this paper define a system to be in equilibrium if the stationary state corresponds to the Boltzmann distribution,
   the relation to other  definitions of equilibrium will be discussed in Sect. \ref{detbal}.
We implement this condition by 
replacing the stationary covariance matrix $E_{ij}^{\circ}$ in the Lyapunov Eq. (\ref{Lyapunov}) by 
the equilibrium covariance matrix $E_{ij}^{\bullet}$ from Eq. (\ref{Ebullet}), by which we  obtain 
\begin{equation}
\label{LyapBoltz0}
2C^\bullet_{ij}/(k_BT)=A_{ik} H_{kj}^{-1} +   A_{jk} H_{ki}^{-1}.
\end{equation}
Inserting the expression for the Langevin matrix $A$ from Eq. (\ref{A}) and using the 
fact that the matrix $U$ is antisymmetric, see Eq. (\ref{U}),  we arrive at  the Lyapunov-Boltzmann equation
\begin{equation}
\label{LyapBoltz}
2C^\bullet_{ij}/(k_BT)=\Gamma_{ik} H_{kj}^{-1} +\Gamma_{jk} H_{ki}^{-1}.
\end{equation}
If for arbitrary Hamiltonian matrix $H$ the  friction matrix $\Gamma$ and the random force correlation
matrix $C$ obey this equation, the Langevin equation given by Eq. (\ref{Langevin}) describes the 
dynamics of an equilibrium system. Conversely, if  $C$ and $\Gamma$
do not satisfy Eq. (\ref{LyapBoltz}) 
the Langevin equation describes a NEQ system. 
Two obvious NEQ scenarios come to mind: 
i) A Newtonian Hamiltonian many-body system coupled to heat baths characterized by different 
temperatures (as will be discussed starting in Section \ref{EntropyProd}),
and ii) a many-body system with off-diagonal friction terms that do not obey 
Eq. (\ref{LyapBoltz}).

Let us give a simple example, 
namely the harmonic oscillator  with Hamiltonian ${\cal H}= K x^2/2 + p^2/(2m)$, which is described
by the Hamiltonian matrix 
 \begin{equation}
H = \begin{pmatrix}
 K & 0  \\
0  & 1/m  \\
 \end{pmatrix}	.
\end{equation}
We choose  a diagonal momentum friction model described by the friction matrix
 \begin{equation}
\Gamma = \begin{pmatrix}
 0 & 0  \\
0  & \gamma/m  \\
 \end{pmatrix}	,
\end{equation}
where the friction force is proportional to the velocity of the  particle and only enters the momentum degree of freedom.
The matrix product appearing on the  right side of the Lyapunov-Boltzmann Eq. (\ref{LyapBoltz}) is given by 
 \begin{equation}
\Gamma H^{-1}  = \begin{pmatrix}
 0 & 0  \\
0  & \gamma  \\
 \end{pmatrix}
\end{equation}
and thus  the equilibrium random  correlation matrix follows as 
 \begin{equation}
 \label{Cbullet}
C^\bullet   = \begin{pmatrix}
 0 & 0  \\
0  & k_BT \gamma  \\
 \end{pmatrix}.
\end{equation}
This  agrees with the well-known result that  the equilibrium Langevin equation of a massive particle 
involves a random field that acts on the momentum degree of freedom only and is proportional 
to the friction coefficient $\gamma$, where the proportionality constant $k_BT$ defines the temperature of the reservoir. 
Conversely, any non-trivial deviation of the matrix $C$ from Eq. (\ref{Cbullet}), i.e. any deviation 
that cannot be
captured by  a modified  temperature,
  indicates a NEQ  system. For the simple
example of a one-dimensional harmonic oscillator 
considered here, this could for example  be the presence of an additional  entry  in  the symmetric matrix $C$,
 e.g.  additional entries in the off-diagonals or in the upper-left diagonal.

%%%%%%%%%%%%%%%%%%%%%%%%%%%%%%%%%%%%%%%%%%%%%%%%%
\subsection{Dynamic approach to the stationary   distribution}
\label{DynApproach}

From the time-dependent analogue   of the Shannon entropy Eq. (\ref{entropy})
\begin{equation}
\label{entropy2}
{\cal S}(t)  /k_B   = -   \int {\rm d} \vec{z}  \rho(\vec{z},t)  \ln \rho(\vec{z},t),
\end{equation}
we obtain by differentiation 
\begin{equation}
\label{entropy2b}
\dot {\cal S}(t)  /k_B   = -   \int {\rm d} \vec{z}  \dot\rho(\vec{z},t) [  \ln \rho(\vec{z},t) +1].
\end{equation}
The time derivative of the density distribution $ \dot\rho(\vec{z},t)$ is determined by the 
Fokker-Planck equation 
\begin{equation}
	\label{FP}
	\dot{\rho}(\vec{z},t)  =\left[   \nabla_k A_{km} z_m +  \nabla_k   \nabla_m  C_{km}\right] \rho(\vec{z},t)
\end{equation}
which follows   via Kramers-Moyal expansion of the Langevin Eq. (\ref{Langevin})  \cite{Risken}.
With the Gaussian  time dependent  distribution Eq. (\ref{rho})
we obtain from Eq. (\ref{FP})   the expression
\begin{eqnarray}
	\label{FP2}
\dot{\rho}(\vec{z},t)  = \rho(\vec{z},t) \times \hspace{1cm}  \\
  \left[  A_{kk} -   A_{km} z_m E^{-1}_{kj} \Delta z_j + 
  C_{ij}(E^{-1}_{ik} \Delta z_k E^{-1}_{jl} \Delta z_l - E^{-1}_{ij})   \right].
\nonumber
\end{eqnarray}
Inserting this into the expression Eq.(\ref{entropy2b}) and calculating all  
 Gaussian expectation values, 
%(see Appendix D) 
we obtain the
final expression for the time derivative of the entropy as 
\begin{equation}
\label{entropy3}
\dot {\cal S}(t)  /k_B   = C_{km} E^{-1}_{km}(t) - A_{kk},
\end{equation}
which holds for equilibrium as well as for NEQ systems.
From the Lyapunov Eq. (\ref{Lyapunov}) we can derive the expression 
\begin{equation}
\label{Lyapunov2}
C_{ij} E^{\circ -1}_{ij}=A_{kk},
\end{equation}
inserting this into Eq.  (\ref{entropy3}) we obtain the alternative expression 
\begin{equation}
\label{entropy4}
\dot {\cal S}(t)  /k_B   = C_{km} [ E^{-1}_{km}(t)  - E^{\circ -1}_{km} ],
\end{equation}
which demonstrates that the entropy change vanishes in the stationary state $ E_{km}(t)  = E^{\circ}_{km}$,
as is expected. We will later come back to Eq. (\ref{entropy4}) as it allows to write down 
one of the  constitutive dynamic equations for NEQ systems.

We next calculate the time derivative of the covariance matrix. 
From  the expression  Eq. (\ref{FP2}) we immediately read off that 
\begin{eqnarray}
\partial \ln {\rho}(\vec{z},t) / \partial t  =  \hspace{1cm}  \\
    A_{kk} -   A_{km} z_m E^{-1}_{kj} \Delta z_j + C_{ij}(E^{-1}_{ik} \Delta z_k E^{-1}_{jl} \Delta z_l - E^{-1}_{ij}),
\nonumber
\end{eqnarray}
which can be rewritten as 
\begin{eqnarray}
\label{deriv1}
\partial \ln {\rho}(\vec{z},t) / \partial t  =  
  \Delta z_i  \Delta z_j  [C_{km}E^{-1}_{kj} E^{-1}_{mi}- A_{ki} E^{-1}_{kj} ]   \hspace{0.3cm}   \\
+ A_{kk}  - C_{km}E^{-1}_{km} -   A_{km} \langle z_m(t) \rangle  E^{-1}_{kj} \Delta z_j .
\nonumber
\end{eqnarray}
On the other hand, using the definition of the Gaussian distribution Eq.(\ref{rho}) we find
\begin{eqnarray}
\frac{ \partial \ln {\rho}(\vec{z},t) }{ \partial t}   =  \frac{\partial }{\partial t} \left\{
- \frac{1}{2}  \ln [ (2\pi)^{2N} \det E(t)] \right. \\
\left. 
-  \frac{1}{2}  [ z_i   - \langle z_i (t) \rangle ] 
  E^{-1}_{ij}(t)   [ z_j   - \langle z_j (t) \rangle]    \right\},
\nonumber
\end{eqnarray}
which can be rewritten as 
\begin{eqnarray}
\frac{ \partial \ln {\rho}(\vec{z},t) }{ \partial t}   =  
 \langle \dot{z}_i(t) \rangle  E^{-1}_{ij} (t) \Delta z_j   \hspace{1cm}   \\
+
\frac{ \dot{E}^{-1}_{kl}  \partial }{\partial E^{-1}_{kl}  } \left\{
 \frac{1}{2}  \ln [ (2\pi)^{-2N} \det E^{-1} (t)] 
 %\right. \\  \left. 
-  \frac{1}{2}  \Delta z_i   E^{-1}_{ij}(t)   \Delta  z_j     \right\} 
\nonumber
\end{eqnarray}
and finally yields 
\begin{equation} \label{deriv2}
\frac{ \partial \ln {\rho}(\vec{z},t) }{ \partial t}   =  
 \langle \dot{z}_i(t) \rangle  E^{-1}_{ij} \Delta z_j  % \hspace{1cm}   \\
+
 \frac{  \dot{E}^{-1}_{kl} }{2}\left\{
 E_{kl}(t)
-   \Delta z_k     \Delta  z_l     \right\} .
\end{equation}
Comparison of Eqs. (\ref{deriv1}) and (\ref{deriv2}) term by term yields
\begin{equation}
\label{invcovT}
\dot {E}^{-1}_{ij}(t)   = - 2 C_{km}  E^{-1}_{kj}(t)  E^{-1}_{mi}(t)  +  
E^{-1}_{jk}(t) A_{ki}  + E^{-1}_{ik}(t) A_{kj}, 
\end{equation}
where we have used  Eq. (\ref{avLangevin}). 
From the basic algebraic relation 
$\dot{E}_{il}(t)  = -    E_{ij}(t)  \dot{E}^{-1}_{jk}(t)  E^{-1}_{kl}(t) $
we finally obtain from Eq.(\ref{invcovT}) the temporal change 
of the covariance matrix  as 
\begin{equation}
\label{covT}
\dot {E}_{ij}(t)   = 2 C_{ij } -  A_{ik} E_{kj}(t) -  A_{jk} E_{ki}(t)
\end{equation}
which,  using Eq.(\ref{Lyapunov}), can be rewritten as 
\begin{equation}
\label{covT2}
\dot {E}_{ij}(t)   =  -  A_{ik}  [ E_{kj}(t)  - E^{\circ}_{kj} ]  -  A_{jk}  [ E_{ki}(t)  - E^{\circ}_{ki} ].
\end{equation}
Note that this expression holds for equilibrium as well as for NEQ  systems.
As would be expected, the temporal change of the covariance matrix vanishes in the stationary state,
i.e. when $ E_{km}(t)  = E^{\circ}_{km}$.

\subsection{Conditions of detailed balance and vanishing probability current} \label{detbal}
Our definition of equilibrium employs the Boltzmann distribution Eq.  (\ref{Boltz})
and not the condition of detailed balance, 
which is often used as the defining property of equilibrium \cite{deGroot}.
The reason for using the Boltzmann condition is that it is very easy to implement, while
the condition of detailed balance is for underdamped  many-particle systems rather involved.
In fact, for  overdamped systems the detailed balance condition becomes equivalent to the condition of 
vanishing probability current \cite{Graham1971,Schnakenberg1976}, which in literature is also called the potential condition. 
We will in this section formulate the  condition for  the probability current to vanish and then compare with the detailed balance condition,
for which the derivation  is presented in Appendix \ref{AppEnew}.

The Fokker-Planck Eq. (\ref{FP})   can be interpreted as a balance equation
\begin{equation}
\label{balance}
	\dot{\rho}(\vec{z},t)  = - \nabla_k {\cal J}_k(\vec{z},t)
\end{equation}
with the probability current $ {\cal J}(\vec{z},t)$ being given as 
\begin{equation}
\label{balance2}
	 {\cal J}_k(\vec{z},t) = - \left[  A_{km} z_m +   \nabla_m  C_{km}\right] \rho(\vec{z},t).
\end{equation}
For the Gaussian distribution  Eq. (\ref{rho}) we obtain (for simplicity we set  $\langle z_i(t) \rangle=0$ here)  for the current
\begin{equation}
\label{balance3}
	 {\cal J}_k(\vec{z},t) = - \left[  A_{km} -    C_{kj} E_{mj}^{-1}   \right] z_m  \rho(\vec{z},t).
\end{equation}
The  probability current vanishes, i.e. $ {\cal J}_k(\vec{z},t)=0$, 
for 
\begin{equation}
\label{balance4}
	 A_{km} E_{mj}=    C_{kj}.
	 \end{equation}
From the equilibrium condition $E_{kj}(t) = E^\bullet_{kj} =  k_BT H^{-1}_{kj}$, Eq. (\ref{Ebullet}),
and the explicit form of the matrix $A$ in Eq. (\ref{A}), we obtain  the
vanishing probability current  condition
\begin{equation}
\label{balance5}
	 \Gamma_{km} H^{-1}_{mj} - U_{kj} =    C_{kj}/(k_BT),
	 \end{equation}
which in the general case is not satisfied  in the equilibrium situation defined by the Lyapunov-Boltzmann Eq. (\ref{LyapBoltz}),
since $C$ is a symmetric matrix, while $U$ is antisymmetric and typically  
$ \Gamma_{km} H^{-1}_{mj}$ is an  asymmetric matrix. 
We conclude that for an underdamped system,   the probability  generally does not vanish, this  is trivially  illustrated by
the fact that a harmonic oscillator performs  orbits in phase space. 
In fact,  current mathematical work is devoted to  separating phase space trajectories of underdamped systems into 
periodic and diffusive parts \cite{Qian2013,Qian2015}. In Appendix \ref{AppCnew}   
we show that in the overdamped (i.e. massless) 
 limit the probability current  in equilibrium  vanishes  if  the friction matrix $\Gamma$ is symmetric.
Since we did not  impose that $\Gamma$ is symmetric so far,
we see that our definition of equilibrium, which is based on the stationary distribution being 
equal to the Boltzmann distribution, is for overdamped systems only equivalent to the vanishing probability current condition for a symmetric friction matrix.
We conclude that  the vanishing probability current condition   is even  in the overdamped limit
 a more restrictive criterion than our Boltzmann distribution criterion.

Furthermore, in Appendix \ref{AppDnew}   we show for the special case of two coupled particles, that the equilibrium 
fluctuation-dissipation theorem holds  if   our Boltzmann definition for  equilibrium  is satisfied and in particular also works
for an asymmetric friction matrix. This suggests that equilibrium definitions based on the Boltzmann distribution and based on the fluctuation-dissipation
theorem are equivalent and that the condition of vanishing probability current  is more restrictive and requires the  symmetry of the friction matrix.

Finally, in  Appendix \ref{AppEnew}  we derive the condition of detailed balance for our underdamped Hamiltonian model
and demonstrate that it is equivalent to the Boltzmann-distribution based criterion for equilibrium only  if the friction matrix is symmetric.
Clearly, an asymmetric friction matrix breaks the physical principle of equal actio and reactio,  so for physical models where
friction is produced e.g. by hydrodynamic interactions,  the friction matrix should be symmetric and our definition of equilibrium 
(based on the Boltzmann distribution) is fully equivalent to the condition of detailed balance. More abstract models 
with asymmetric friction matrices are conceivable, for such models the distribution is predicted to be of the  Boltzmann type 
if the Lyapunov-Boltzmann Eq. (\ref{LyapBoltz}) is satisfied, yet the condition of detailed balance is violated. 

\subsection{Time-dependent free energy: extremal and stability properties}

The time derivative of the free energy expression Eq. (\ref{freeen}) reads
\begin{equation}
\label{freeenderiv}
{\cal \dot{F}} (t)   =   H_{ij}   \langle \dot{z}_i(t)  \rangle  \langle   z_j(t)  \rangle  + H_{ij} \dot{E}_{ij}(t)  /2   - k_B T {\cal \dot{S}} (t) 
\end{equation}
and using Eqs. (\ref{avLangevin}), (\ref{entropy3}), (\ref{covT}) is explicitly given by
\begin{eqnarray}
\label{freeenderiv2}
{\cal \dot{F}} (t)   =   -\langle z_j (t)  \rangle H_{ji}  A_{ik}    \langle   z_k(t)  \rangle  + 
H_{ij}C_{ij } -  H_{ij} A_{ik} E_{kj}(t) \nonumber  \\
- k_BT C_{km} E^{-1}_{km}(t) +k_BT  A_{kk}.  \hspace{1cm}
\end{eqnarray}
After some  manipulation this expression can be rewritten as 
\begin{eqnarray}
\label{freeenderiv3}
{\cal \dot{F}} (t)   =   -\langle z_j (t)  \rangle H_{ji}  \Gamma_{ik}    \langle   z_k(t)  \rangle   \hspace{1cm}   \\
-  C_{km} [ H_{ml} - k_BT  E^{-1}_{ml}(t) ]\frac{E_{lj}(t)}{k_BT}    [ H_{jk} - k_BT  E^{-1}_{jk}(t) ]  \nonumber.
\end{eqnarray}
The quadratic form of Eq. (\ref{freeenderiv3}) directly demonstrates that 
for the equilibrium distribution, defined by  $E_{kj}(t) = E^\bullet_{kj} =  k_BT H^{-1}_{kj}$  and $   \langle z_k(t)  \rangle =0$,
the free energy is stationary and does not change in time, i.e. ${\cal \dot{F}} (t)   = 0$.
This is somewhat trivial since this just reflects  that the equilibrium distribution is a special case of a stationary distribution.
More importantly, 
from the fact that $H$, $C$ and $E$ are symmetric matrices that are positive-definite or
 semi-positive-definite, we 
derive  in Appendix \ref{AppC} that the free energy does not increase in time, i.e. 
\begin{equation}
\label{Fdotminus}
{\cal \dot{F}} (t)   \leq 0
\end{equation}
in all generality,
this means that the equilibrium distribution  is stable with respect to perturbations. 
For this we have used  that the first term of Eq. (\ref{freeenderiv3}) 
is not negative since the product of two positive semi-definite matrices is also 
positive semi-definite.
Equation (\ref{Fdotminus})  corresponds to the second law of thermodynamics, here derived for 
many-body systems described by quadratic Hamiltonians and  Langevin equations 
with friction terms and random fields. These results are graphically 
illustrated in Fig. 1: Figure 1a)  shows that  the free energy ${\cal F} (t) $ is minimal in  equilibrium 
  defined by  $E_{kj}(t) = E^\bullet_{kj} $  and $   \langle z_k(t)  \rangle =0$, 
as demonstrated by Eq. (\ref{freeen2}). 
Figure 1b) shows that the time derivative of the free energy ${\cal \dot{F}} (t)$
is negative except in equilibrium  where it vanishes, as follows from  Eq. (\ref{Fdotminus}).
This indicates that a non-stationary  distribution
flows monotonically towards the equilibrium stationary distribution, as indicated by the red arrows in the top graph.

We finally  derive a set of constitutive dynamic equations, which we will in the next section
generalize for NEQ systems. For an equilibrium system the Lyapunov-Boltzmann condition 
Eq. (\ref{LyapBoltz}) is satisfied and the stationary covariance matrix $E^\circ_{kj} $ is given by
the equilibrium state, i.e.  $E^\circ_{kj} = E^\bullet_{kj} =  k_BT H^{-1}_{kj}$.
Comparison of the expression  for the   entropy production Eq. (\ref{entropy4}) and the
derivative of the free energy Eq. (\ref{extremize}) yields
\begin{equation}
\label{const1}
\dot{\cal S}(t)   = -   \frac{2}{k_BT}C_{ij}  \frac{\partial   {\cal F} (t)}{\partial  E_{ij}},
\end{equation}
which is a  simple relation between the   entropy production $\dot{\cal S}$ and the
free energy derivative, the first constitutive dynamic relation. The other dynamic relations are obtained
from  the time derivative of the covariance matrix Eq. (\ref{covT2}),
\begin{equation}
\label{const2}
\dot {E}_{ij}(t)   = -  \frac{2}{k_BT}\left[  
A_{ik} E^\bullet_{kl}  \frac{\partial   {\cal F} (t)}{\partial  E_{lm} } E_{jm} +
A_{jk} E^\bullet_{kl}  \frac{\partial   {\cal F} (t)}{\partial  E_{lm} } E_{mi}
\right],
\end{equation}
and by comparing  Eq.  (\ref{avLangevin}) with the derivative of the free energy Eq. (\ref{freeenb}) 
with respect to $\langle z_i(t) \rangle$,
\begin{equation}
\label{const3}
\langle \dot {z}_i(t) \rangle    = -  \frac{1}{k_BT}  
A_{ik} E^\bullet_{kl}  \frac{\partial   {\cal F} (t)}{\partial  \langle z_l \rangle } .
\end{equation}
Equations (\ref{const1}), (\ref{const2}) and (\ref{const3}) are the constitutive  dynamic equations
for equilibrium systems
that relate temporal changes of all relevant quantities to derivatives of the free energy with respect
to the state variables, i.e. to generalized thermodynamic forces. 

\begin{figure*}	
	\centering
	\includegraphics[width=12cm]{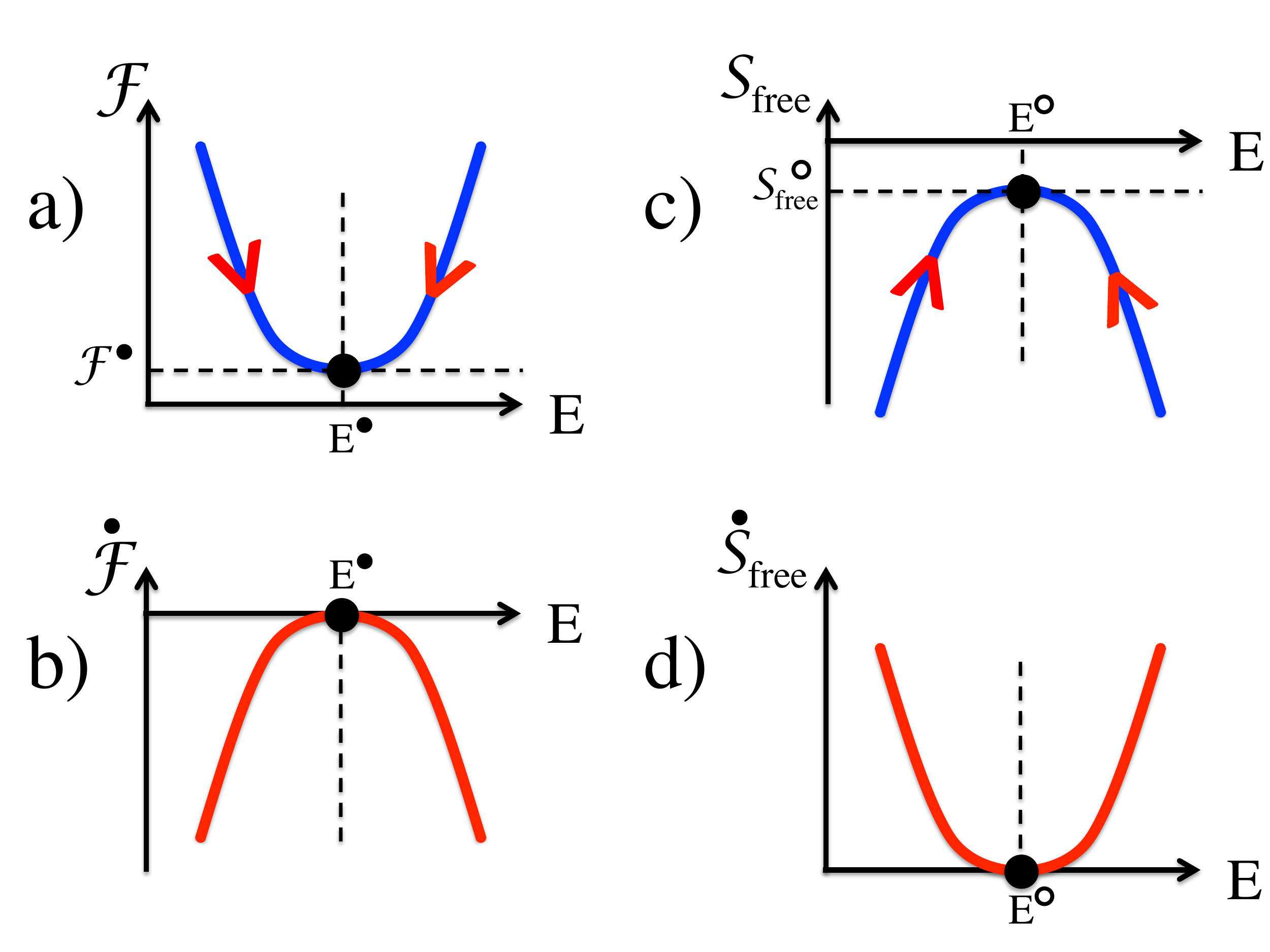}
\caption{ Graphical  illustration of extremal and stability properties of equilibrium and non-equilibrium 
(NEQ) systems.
For an equilibrium system, the free energy in a) takes its minimal value ${\cal F}^\bullet$  given by Eq. (\ref{Fbullet})
at the equilibrium value of the  covariance matrix $E^\bullet$, determined  by Eq.  (\ref{Ebullet}). The free energy production 
$\dot{\cal F}$ in b) is negative and vanishes at the equilibrium state, see Eq. (\ref{Fdotminus}).
 This means that the system is stable
with respect to perturbations and flows towards the equilibrium state, as indicated by red arrows in a).
For a NEQ  system, the   free entropy 
$ {\cal S}_{\rm free}(t) $  in c) 
takes its maximal value ${\cal S}_{\rm free}^\circ$
at the stationary  value of the  covariance matrix $E^\circ$. The   free entropy production 
$ \dot{\cal S}_{\rm free}(t)$  in d) is positive  and vanishes in  the stationary  state. 
This means that the system is stable
with respect to perturbations and flows towards the NEQ stationary state, as indicated by red arrows in c).
		}
	\label{fig1}
\end{figure*}

\section{Truly Non-Equilibrium Systems}

\subsection{Extremal and stability properties of the free entropy}
\label{extremalprop}

We remark  that  the word non-equilibrium  (NEQ) typically refers to two very different situations: 
For  a NEQ  system, i.e. when  the Lyapunov-Boltzmann condition 
Eq. (\ref{LyapBoltz}) is not satisfied,  the stationary distribution  is a NEQ  stationary distribution,
which is characterized by a non-vanishing positive entropy production. For such 
a system an equilibrium distribution does not exist.
An equilibrium system is one where the Lyapunov-Boltzmann condition Eq. (\ref{LyapBoltz}) is  satisfied,
but even such a system is off equilibrium  as long as it has not settled in its stationary equilibrium distribution.

Obviously, it is not possible to use  the
constitutive equilibrium dynamic equations  (\ref{const1}), (\ref{const2}) and (\ref{const3}) 
 in the NEQ  case,   because of the appearance of  
the equilibrium  temperature $T$. The temperature can be trivially
eliminated by introducing a modified thermodynamic potential, which is called the
free entropy and which is, for the equilibrium scenario  using Eqs.  (\ref{internalenergy}),
(\ref{freeen}) and (\ref{Ebullet}),    given by 
\begin{equation}
\label{Sfree1}
\frac{{\cal S}_{\rm free}(t)}{k_B} \equiv  
-  \frac{{\cal F} (t)}{k_B T} = \frac{{\cal S}(t) }{k_B}
 -  \frac{E^{\bullet -1}_{ij} E_{ij}(t) }{2} 
-\frac{E^{\bullet -1}_{ij}   \langle z_i(t)  \rangle  \langle   z_j(t) \rangle }{2}.
\end{equation}
As mentioned before, the  equilibrium  
free entropy had been originally introduced  by Massieu in 1869 \cite{Callen,Balian}
 and the advantage of this functional was  already recognized by Planck \cite{Planck} and 
 Schrödinger \cite{Schroedinger}. The free entropy  concept 
 is particularly useful in the current NEQ setting, since for the NEQ systems we
 will consider  there is no unique  temperature.   

In fact, the free entropy concept allows for straightforward generalization to the NEQ  case.
For this we replace  $E^{\bullet -1}_{ij} $ in Eq. (\ref{Sfree1}) by $E^{\circ -1}_{ij} $,
after which we obtain the NEQ  version of the free entropy
\begin{equation}
\label{Sfree3}
\frac{ {\cal S}_{\rm free}(t)}{k_B}   =  \frac{{\cal S}(t)}{k_B} -  \frac{ E^{\circ -1}_{ij}  E_{ij}(t)  }{2}  
- \frac{ E^{\circ -1}_{ij}    \langle z_i(t)  \rangle  \langle   z_j(t)\rangle  }{2}.
%+ \frac{  t {\cal \dot{S}}^\circ_{\rm res}}{k_B}
\end{equation}
%
%and where we have added a term proportional to $ \cal{\dot{S}}^\circ_{\rm res}$,
Note that  the entropy production in the stationary NEQ  state, which 
for suitably chosen friction and random strength matrices can be described as being 
due to heat fluxes in and out of  heat reservoirs,  is, as illustrated  in Eq. (\ref{intro1}), not included in the free entropy,
 but will  be discussed in Section \ref{EntropyProd}.
  Clearly, for an equilibrium system, 
for which $ E^{\circ -1}_{ij}= E^{\bullet -1}_{ij}= H_{ij} /(k_BT)$,
the general expression Eq. (\ref{Sfree3}) reduces to the equilibrium expression Eq. (\ref{Sfree1}).
Similar  functionals, which in a NEQ stationary state are extremal, were derived previously 
\cite{Graham1971,Schloegl1971,Procaccia1976,Schnakenberg1976,Kwon2005,Esposito2007,Ge2009,Ge2010,Qian2013,Qian2015},  
in Appendix \ref{AppGnew} we show that the free entropy expression Eq.(\ref{Sfree3}) is 
equivalent to  the Kullback-Leibler entropy \cite{Kullback1951,Schloegl1971}.
 Our model differs from previous approaches in that the definition of separate friction and random strength matrices
 will allow us to describe the  coupling to multiple heat baths with different  well-defined temperatures.

Using the free entropy, 
the constitutive dynamic equations (\ref{const1}), (\ref{const2}) and (\ref{const3})  
 can be rewritten as
\begin{equation}
\label{const1b}
\dot{\cal S}(t)   =2C_{ij}\frac{\partial   {\cal S}_{\rm free} (t) /k_B }{\partial  E_{ij}},
\end{equation}
\begin{eqnarray}
\label{const2b}
\dot {E}_{ij}(t)   =&
2 A_{ik} E^\circ_{kl}  \frac{\partial   {\cal S}_{\rm free} (t) /k_B  }{\partial  E_{lm} } E_{jm} \nonumber \\
&+ 2 A_{jk} E^\circ_{kl}  \frac{\partial   {\cal S}_{\rm free} (t) /k_B   }{\partial  E_{lm} } E_{mi} ,
\end{eqnarray}
\begin{equation}
\label{const3b}
\langle \dot {z}_i(t) \rangle    = 
A_{ik} E^\circ_{kl}  \frac{\partial   {\cal S}_{\rm free} (t) /k_B   }{\partial  \langle z_l \rangle } ,
\end{equation}
where the temperature has obviously (and trivially) disappeared.
When Eq.(\ref{Sfree3})  is used in conjunction with  the constitutive dynamic equations
 (\ref{const1b}), (\ref{const2b}) and (\ref{const3b}), the exact dynamic evolution equations 
 (derived for general NEQ  systems) Eqs. (\ref{entropy4}), (\ref{covT2}) and (\ref{avLangevin}) 
are reproduced, this independently confirms the validity of   the expression Eq. (\ref{Sfree3}).

We next show that  the free entropy has very similar properties for NEQ  systems
as the free energy has
for equilibrium systems, namely ${\cal S}_{\rm free}(t)$ is extremal and in fact maximal in the 
stationary NEQ 
state and the stationary state is stable in the sense that $\dot{\cal S}_{\rm free}(t) \geq 0$.
For this we basically repeat the steps leading to the minimal condition of the free energy Eq. (\ref{freeen2}).
The extremum of  $  {\cal S}_{\rm free}(t) $
 is  given  by  $\langle \vec{z}(t)   \rangle=0$  and determined by 
\begin{equation}
	\label{extremizeS}
\frac{  \partial  {\cal S}_{\rm free} (t) /k_B  }{\partial E_{ij}}   =  
E^{-1}_{ji}(t) /2 -  E^{\circ -1}_{ji} /2   =0,
\end{equation}
the solution of which yields the  time-independent stationary  distribution,
i.e. $E_{ij}(t) =  E^{\circ}_{ij}$.
 With the stationary covariance matrix $E^{\circ}_{ij}$
 many observables can be calculated, for example the 
   internal  energy
follows  according to  Eq. (\ref{internalenergy}).
The stationary   free entropy  is given by 
\begin{equation}
 {\cal S}^{\circ}_{\rm free} /k_B    =   (1/2) \ln [ (2\pi )^{2N}\det E^\circ ].
\end{equation}
The second derivative of the  free entropy is given by
\begin{equation}
\frac{ \partial^2 {\cal S}_{\rm free} (t) /k_B }{   \partial E_{kl}   \partial E_{ij}  }   =  
- E^{-1}_{ik}(t) E^{-1}_{lj}(t)      /2.
\end{equation}
Around the stationary  state   $E_{ij} = E^{\circ}_{ij}$ 
 the   free entropy thus  is to second order  given by
 \begin{eqnarray} \label{Ssecond}
&  {\cal S}_{\rm free} (t) /k_B -  {\cal S}^\circ_{\rm free} /k_B   \simeq   \\
 & E^{\circ -1}_{ik} E^{\circ -1}_{lj}(E_{ij}(t)- E^{\circ}_{ij } k_BT)  (E_{kl}(t)-E^{\circ}_{kl} k_BT) /2 , \nonumber
\end{eqnarray}
which  is positive
since $E^{\circ -1} $ is symmetric and positive definite 
(the general proof for this is given in Appendix \ref{AppC}).

The  free entropy production follows from Eq. (\ref{Sfree3}) by taking a time derivative  as 
\begin{equation}
\label{Sfree4}
\frac{ \dot{\cal S}_{\rm free}(t)}{k_B}   =  \frac{\dot{\cal S}(t)}{k_B} -  \frac{ E^{\circ -1}_{ij}  \dot E_{ij}(t)  }{2}  
-   E^{\circ -1}_{ij}    \langle \dot z_i(t)  \rangle  \langle   z_j(t)\rangle.
\end{equation}
Using our previous results for $\dot{\cal S}(t)$, Eq. (\ref{entropy4}), 
for $ \dot E_{ij}(t)$, Eq. (\ref{covT2}), and for $  \langle \dot z_i(t)  \rangle$, Eq. (\ref{avLangevin}), we arrive 
after a few intermediate steps at
\begin{eqnarray}
\label{Sfree5}
{\cal \dot{S}}_{\rm free} (t)/ k_B    =\langle z_j (t)  \rangle E^{\circ -1}_{ji}  A_{ik}    \langle   z_k(t)  \rangle   \hspace{1cm}   \\
+  C_{km} [ E^{\circ -1}_{ml} - E^{-1}_{ml}(t)]E_{lj}(t)     [ E^{\circ -1}_{jk} - E^{-1}_{jk}(t) ]
     \nonumber.
\end{eqnarray}
The quadratic form of this expression  shows that 
in the stationary state, defined by  $E_{kj}(t) = E^\circ_{kj} $  
and $   \langle z_k(t)  \rangle =0$,
the free entropy production of the system vanishes. 
More importantly, 
from the fact that $E^\circ$, $C$ and $E$ are symmetric matrices 
 that are semi-positive-definite or positive-definite, it 
follows that the   free entropy does not decrease in time, i.e. 
\begin{equation}
\label{Sdotpositive}
 {\cal \dot{S}}_{\rm free} (t)    \geq 0
\end{equation}
in all generality,
this means that the stationary NEQ  distribution  is stable with respect to perturbations,
the  proof  is given in Appendix \ref{AppC}.
In writing Eq. (\ref{Sdotpositive}) we have used that the first term of Eq. (\ref{Sfree5}) 
is not negative since the product of 
the  two positive-definite matrices $A$ and $E^\circ$  is also 
positive-definite,   \textcolor{red}{ where  positive definiteness of the 
asymmetric matrix $A$ is equivalent to demanding that all eigenvalues of the symmetric part of
$A$  are positive, which is more restrictive than the condition that the eigenvalues of $A$
have all positive real components, 
as required for the existence of a stationary state in  Sect. \ref{stationary}.}
This result  is  here derived for interacting
many body systems described by quadratic Hamiltonians and is  graphically 
illustrated in Fig. 1. Figure 1c)  shows that  the   free entropy  
$ {\cal S}_{\rm free} (t) $ is maximal at the stationary 
distribution   defined by  $E_{kj}(t) = E^\circ_{kj} $  and $   \langle z_k(t)  \rangle =0$, 
which follows from Eq. (\ref{Ssecond}). 
Figure 1d) shows that the time derivative of the  free entropy $ {\cal \dot{S}}_{\rm free} (t)$
is positive except at the stationary distribution  where it vanishes, as follows from  Eq. (\ref{Sfree5}).
This indicates that the system  flows monotonically towards the stationary  state, 
as indicated by the red arrows in Fig. 1c).

The free entropy maximization principle  applies to 
NEQ  and equilibrium systems alike.
On the one hand this allows to treat NEQ and equilibrium systems within a unified framework
and thereby eliminates a disturbing schism in the description  of these systems.
On the other hand, the usage of the free entropy, 
which does not include the stationary  reservoir entropy production (which 
is related to the so-called  house-keeping entropy  \cite{Oono1998,Hatano2001,Speck2005} and
 increases linearly
in time in a stationary NEQ  state,  see Eq. (\ref{intro1}) and as discussed  in the next section),
brings   a significant  methodological advantage over 
early  approaches to NEQ systems,
 according to which  stationary NEQ states are defined by an extremum
of the total entropy production (including the reservoirs)
and stability criteria  invoke the time derivative of the entropy production, 
i.e. the second time derivative of the total entropy \cite{deGroot}.
The connection between the free entropy  production (excluding the reservoirs) 
and the total entropy production (including the reservoirs) is  discussed  
in the Conclusions and also  in Appendix \ref{AppE}.

\subsection{Stationary entropy production for particles coupled to temperature reservoirs}
\label{EntropyProd}

The calculations so far were completely general and no restrictions on the  type of the 
Hamiltonian matrix $H$, the friction matrix $\Gamma$ and the random correlation matrix $C$ were imposed. 
To gain insight into the  entropy production of the reservoirs,
denoted by $\dot{\cal S}_{\rm res}(t)$, we need to restrict the discussion to Newtonian systems, meaning that 
in the Hamiltonian the spatial and momentum coordinates decouple and  the kinetic energy is diagonal. 
 This will allow  to ascribe well-defined temperatures to different heat baths,
which will then be used to calculate the reservoir entropy production from the individual heat fluxes  between the system and the 
heat baths.
To conveniently use the  symmetry of such Newtonian systems, we 
switch to particle indices,  denoted by greek symbols,   with which 
the Hamiltonian can be written as 
\begin{equation}
 {\cal H} (\vec{y})   =  y_\alpha H_{\alpha \epsilon}  y_\epsilon /2.
\end{equation}
Here  we introduced the particle state vector
 \begin{equation}
y_\alpha(t) = (x_\alpha(t), p_\alpha(t))^T
\end{equation}
where $x_\alpha(t)$ and $p_\alpha(t)$ are the position and the momentum of particle $\alpha$. 
The $N\times N$ entries of the Hamiltonian matrix $H_{\alpha \epsilon}$ each consist
of $2 \times 2$ matrices. These sub matrices are expanded in terms of the 
matrices
 \begin{equation} \label{2by2}
 u  = \begin{pmatrix}
 0 & 1   \\
-1  & 0   \\
\end{pmatrix},   \,\, \,\,
 s  = \begin{pmatrix}
 1 & 0   \\
0  & 0   \\
\end{pmatrix},   \,\, \,\,
 r  = \begin{pmatrix}
 0 & 0   \\
0  & 1   \\
\end{pmatrix}.
\end{equation}
From the matrix products
\begin{equation} \label{2by2b}
 us=ru  = \begin{pmatrix}
 0 & 0   \\
-1  & 0   \\
\end{pmatrix},   \,\, \,\,
 ur=su  = \begin{pmatrix}
 0 & 1   \\
0  & 0   \\
\end{pmatrix}
\end{equation}
it is easily seen that all 
$2 \times 2$ matrices can be expanded in terms of the four matrices 
$s$, $r$, $us$ and $ur$, which thus form a convenient complete set.
We will make in the following repeated use of the properties
\begin{equation} \label{2by2c}
s^2=s,    \,\, \,\,
r^2=r,    \,\, \,\,
u^2=-\mathbb{1},    \,\, \,\,
rs= \mathbb{0}=sr.
\end{equation}
The Langevin Eq. (\ref{Langevin})  can be written as
\begin{equation}
	\label{Langevin2}
	\dot{y}_\alpha (t)  = -A_{\alpha \epsilon} y_\epsilon(t) +\Phi_{\alpha \epsilon }F_\epsilon  (t).
\end{equation}
where  $A_{\alpha \epsilon}$  is  given by
\begin{equation}
	\label{A2}
	A_{\alpha \epsilon}   =  - U_{\alpha \gamma}  H_{\gamma \epsilon}   +  \Gamma_{\alpha \epsilon}
\end{equation}
and  $U_{\alpha \gamma} $    can be written as  $ U_{\alpha \gamma} =    u \delta_{\alpha \gamma}$.
The Hamiltonian matrix is for Newtonian systems given by
\begin{equation}
	\label{Ham2}
	H_{\alpha' \epsilon}   =   s h_{\alpha' \epsilon}  +  r \delta_{\alpha' \epsilon} /m_{\alpha'}
	\end{equation}
where $h_{\alpha \epsilon}$ is a general  $N\times N$ symmetric
 interaction matrix that only acts on positional degrees of freedom
(and thus is multiplied by the matrix $s$) and the second term is the kinetic energy 
which is diagonal in the momentum degrees of freedom (and thus is multiplied by the matrix $r$)
and $m_{\alpha'}$ is the mass of particle $\alpha'$. 
Note that the primed index $\alpha'$ is not summed over.
From the product properties of the 
$2 \times 2$ matrices 
the inverse Hamiltonian matrix follows as 
\begin{equation}
	\label{Ham3}
	H^{-1}_{\alpha' \epsilon}   =   s h^{-1}_{\alpha' \epsilon}  +  r \delta_{\alpha' \epsilon} m_{\alpha'}.
	\end{equation}
To allow for a clear definition of  reservoir temperatures, 
 we will in the remainder  treat momentum-diagonal friction, 
 for which the friction matrix $\Gamma$  is reduced to the momenta entries as 
\begin{equation} \label{Gamma}
 \Gamma_{\alpha' \epsilon} = r \gamma_{\alpha' \epsilon}   / m_{\alpha'}
\end{equation}
and where the matrix  $ \gamma_{\alpha' \epsilon}$  is diagonal and given 
by $ \gamma_{\alpha' \epsilon} =  \delta_{\alpha' \epsilon}  \gamma_{\alpha'}$,
where $ \gamma_{\alpha'}$ is the friction coefficient of particle $\alpha'$. 
The Lyapunov-Boltzmann  condition  Eq. (\ref{LyapBoltz})
in terms of particle indices reads 
\begin{equation}
\label{LyapBoltz2}
    2 C^\bullet_{\alpha \epsilon }/(k_BT)=\Gamma_{\alpha \gamma } H_{\gamma \epsilon}^{-1} +  
     \Gamma_{\epsilon \gamma } H_{\gamma \alpha}^{-1},
\end{equation}
which, using Eqs. (\ref{Ham3}) and  (\ref{Gamma}) and the matrix product properties Eq. (\ref{2by2c}),
leads to 
\begin{equation}
\label{LyapBoltz3}
     C^\bullet_{\alpha' \epsilon }  =  r   \delta_{\alpha' \epsilon} \gamma_{\alpha'}  k_BT, \hspace{0.5cm}
     \Phi^\bullet_{\alpha' \epsilon }  = r    \phi^\bullet_{\alpha' \epsilon }=
      r   \delta_{\alpha' \epsilon}   \sqrt{\gamma_{\alpha'} k_BT}.
\end{equation}
These expressions show that the random correlation and random strength matrices
are  diagonal in the particle indices and
proportional to $r$  and thus only couple momentum degrees to each other.
The NEQ generalization of Eq. (\ref{LyapBoltz3}) 
for reservoirs with different temperatures reads 
\begin{equation}
\label{LyapBoltz4}
     C_{\alpha' \epsilon }  =  r \gamma_{\alpha'}  \delta_{\alpha' \epsilon} /\beta_{\alpha'}, \hspace{1cm}
     \Phi_{\alpha' \epsilon }  =  r   \delta_{\alpha' \epsilon} \sqrt{\gamma_{\alpha'} /\beta_{\alpha'}},
\end{equation}
where $\beta_{\alpha'}=1/(k_B T_{\alpha'})$ denote the different inverse 
 thermal energies of 
reservoirs, each characterized  by a temperature $T_{\alpha'}$.
The expressions Eq. (\ref{LyapBoltz4}) are central to our paper as they define the NEQ 
model we are using to derive all following results.

To calculate an explicit expression for the reservoir entropy production we 
multiply the Langevin equation
(\ref{Langevin2}) for the momentum component $p_\alpha(t)$ 
by  $p_\alpha(t) $ and use Eqs. (\ref{A2}) and (\ref{Ham2}) to obtain 
\begin{eqnarray}
	\label{Sprod1}
p_{\alpha'}(t)  \dot{p}_{\alpha'} (t)  =  d(   p^2_{\alpha'} (t))/(2dt) = \hspace{3.9cm}  \\
  - p_{\alpha'}(t)   h_{\alpha' \epsilon}  x_\epsilon(t)  
 - p_{\alpha'}(t)     \gamma_{\alpha'} p_{\alpha'}(t)/m_{\alpha'} 
+ p_{\alpha'}(t)  \Phi_{\alpha' \alpha'} F_{\alpha'}(t), \nonumber
\end{eqnarray}
where we again note  that  the primed index $\alpha'$ is not summed over.
From this expression the average heating rate of the system due to  reservoir $\alpha'$,
i.e. the work performed  by the random force per unit time, 
the last term in Eq. (\ref{Sprod1}),  minus the friction work
dissipated  by the particle $\alpha'$ per unit time, the second-last term in Eq. (\ref{Sprod1}),  
 turns out to be
\begin{eqnarray}
	\label{Sprod2}
\dot Q_{\alpha'}(t) &=  \Phi_{\alpha' \alpha'}   \langle p_{\alpha'}(t)  F_{\alpha'}(t)  \rangle /m_{\alpha'}
-  \gamma_{\alpha'}  \langle p_{\alpha'}(t)     p_{\alpha'}(t) \rangle  /m^2_{\alpha'}  \nonumber  \\
& = \frac{1 }{2m_{\alpha' }   } \frac{ {\rm d}   \langle  p^2_{\alpha'} (t)  \rangle  }{ {\rm d}t}
+   h_{\alpha' \epsilon}  \langle  p_{\alpha'}(t)    x_\epsilon(t)   \rangle /m_{\alpha'}. % \hspace{1.9cm}
\end{eqnarray}
Since the last line of Eq. (\ref{Sprod2}) is nothing but the time derivative 
of the sum of the kinetic and potential energies  of particle $\alpha'$,
we see that the reservoir heating rate $\dot Q_{\alpha'}(t)$
balances the particle energy at each instance of time
(a clear consequence of the quadratic Hamiltonian approximation). 
For the  entropy production  of all reservoirs $\dot{\cal S}_{\rm res}(t)$ we  thus
 obtain from Eq. (\ref{Sprod2}) the expression
\begin{equation}
	\label{Sprod5}
\frac{ \dot{\cal S}_{\rm res}(t) }{k_B} \equiv    - \beta_\alpha    \dot Q_{\alpha}(t)
 = -  \frac{ \beta_\alpha  }{2m_{\alpha}}    \frac{ {\rm d}    \langle p^2_{\alpha} (t) \rangle }{ {\rm d} t}
 -  \frac{ \beta_\alpha h_{\alpha \epsilon}  }{ m_{\alpha}} \langle  p_\alpha x_\epsilon  \rangle.
 \end{equation}
We now replace momenta by velocities 
$v_{\alpha'}(t) = p_{\alpha'}(t) / m_{\alpha'}$.
From the fact that in the stationary state 
\begin{equation}
\frac{\rm d}{{\rm d} t} \langle x_\alpha  x_\epsilon \rangle^\circ 
=\langle x_\alpha  v_\epsilon  \rangle^\circ  +\langle v_\alpha  x_\epsilon  \rangle^\circ  = 0
\end{equation}
and $ {\rm d}  \langle v^2_\alpha  \rangle^\circ /   {\rm d} t= 0$
and using that $h_{\alpha \epsilon}$ is symmetric, we obtain for the 
reservoir entropy production in the stationary state
\begin{equation}
	\label{Sprod6}
\dot{\cal S}_{\rm res}^\circ / k_B  = -
 \beta_\alpha   h_{\alpha \epsilon}    \langle x_\epsilon v_\alpha \rangle^\circ =
\frac{1}{2}
 h_{\alpha \epsilon}\langle x_\epsilon v_\alpha \rangle^\circ (  \beta_\epsilon - \beta_\alpha ).
\end{equation}
This expression shows that one  necessary condition for  a non-zero 
 stationary reservoir entropy production 
is that reservoirs have different temperatures. Other necessary conditions are a non-vanishing 
stationary position-velocity coupling $\langle x_\epsilon v_\alpha \rangle^\circ$,
which for Newtonian Hamiltonian systems is only obtained off equilibrium,
 and a non-vanishing  interaction strength $ h_{\alpha \epsilon}$.
To obtain explicit  results for the stationary reservoir entropy production we need
to calculate the  position-velocity correlations $\langle x_\epsilon v_\alpha \rangle^\circ$,
which  do not vanish even in the overdamped massless limit for a NEQ system. 
For this  a systematic perturbative scheme is introduced in the next section.

%%%%%%%%%%%%%%%%%%%%%%%%%%%%%%%%%%%%%%%%%%%%%%%%
\subsection{Perturbative solution of  the Lyapunov equation}

The solution of the Lyapunov equation (\ref{Lyapunov})
for $N$ particles described by a state vector with $2N$ components  consists of
determining all $2N(2N+1)/2$  entries of the symmetric covariance matrix $E$. The calculation
is cumbersome even for only $N=2$ particles \cite{Netz2018}.
Here we introduce a  systematic expansion of the Lyapunov equation 
with the particles mass as the perturbation parameter, which employs  a
projection  of the matrix equations
onto the  complete set of $2 \times 2$ matrices  Eqs.  (\ref{2by2}) and (\ref{2by2b}) introduced in the previous section.
This expansion is subtle, since the leading-order result for the covariance matrix in the limit $m_\alpha \rightarrow 0 $ 
is not obtained by   taking this limit upfront in the Langevin equation.
In fact, the overdamped limit is  commonly obtained by setting 
$m_\alpha = 0 $  in the Langevin equation \cite{Risken}, which correctly describes the long-time particle dynamics but obviously misses 
the short-time ballistic particle dynamics and leads to divergent instantaneous particle velocities. Since we need 
position-velocity correlations in order to estimate the entropy production according to Eq. (\ref{Sprod6}), it is advisable 
to keep velocities to leading order as   $m_\alpha \rightarrow 0$.
As turns out, position-velocity correlations that result from the perturbation calculation
stay finite even in the $m_\alpha \rightarrow 0 $ limit.

To proceed, we expand the covariance matrix as
\begin{equation}
	\label{Sprod3}
E_{\alpha \beta} =\langle x_{\alpha}x_{\beta}\rangle s +\langle p_{\alpha}p_{\beta}\rangle  r + 
\langle x_{\alpha}p_{\beta}\rangle  ut +\langle p_{\alpha}x_{\beta} \rangle  us 
\end{equation}
and  insert the expressions for $A$, Eq. (\ref{A2}),
$H$, Eq. 	(\ref{Ham2}),   $\Gamma$,  Eq.   (\ref{Gamma}),  $C$,  Eq. (\ref{LyapBoltz4}) and $E$, Eq. (\ref{Sprod3}),
 into the Lyapunov Eq. (\ref{Lyapunov}).
The resulting expression splits  into an equation  proportional to $s$ for $\alpha \neq \beta$
\begin{equation}
	\label{Lyaps1}
0=\langle x_\alpha v_\beta \rangle^\circ +\langle x_\beta v_\alpha \rangle^\circ, 
\end{equation}
an equation  proportional to $s$ for $\alpha =  \beta$
\begin{equation}
	\label{Lyaps2}
0=\langle x_\alpha v_\alpha \rangle^\circ, 
\end{equation}
an equation  proportional to $r$ for $\alpha' \neq \beta'$
\begin{eqnarray}
	\label{Lyapt1}
0= 
(\gamma_{\alpha'} m_{\beta'} +\gamma_{\beta'} m_{\alpha'}) \langle v_{\alpha'} v_{\beta'} \rangle^\circ \\
+ m_{\beta'}  h_{\alpha' \gamma}\langle x_\gamma v_{\beta'} \rangle^\circ
+ m_{\alpha'} h_{\beta' \gamma}\langle x_\gamma v_{\alpha'} \rangle^\circ, \nonumber
\end{eqnarray}
an equation  proportional to $r$ for $\alpha' = \beta'$
\begin{equation}
	\label{Lyapt2}
1/  \beta_{\alpha'} = 
 m_{\alpha'} \langle v_{\alpha'} v_{\alpha'} \rangle^\circ 
+ \boxed{ m_{\alpha'} h_{\alpha' \gamma}  \langle x_\gamma v_{\beta'} \rangle^\circ /\gamma_{\alpha'}} ,
\end{equation}
an equation  proportional to $ur$  for $\alpha' \neq \beta'$
\begin{equation}
	\label{Lyapur1}
\boxed{ m_{\beta'} \langle v_{\alpha'} v_{\beta'} \rangle^\circ}= 
\gamma_{\beta'} \langle x_{\alpha'} v_{\beta'} \rangle^\circ
+h_{\beta' \gamma} \langle x_\gamma x_{\alpha'} \rangle^\circ,
\end{equation}
and an equation  proportional to $ur$  for $\alpha' = \beta'$
\begin{equation}
	\label{Lyapur2}
m_{\alpha'} \langle v_{\alpha'} v_{\alpha'} \rangle^\circ=
h_{\alpha' \gamma} \langle x_\gamma x_{\alpha'} \rangle^\circ,
\end{equation}
where we have converted all momenta $p_{\alpha'}$ to velocities 
$v_{\alpha'}= p_{\alpha'} / m_{\alpha'}$. 
Note that there are also two equations proportional to $us$ which however are equivalent to the
equations proportional to $ur$. 
The limit $m_{\alpha'} \rightarrow 0$ must be taken with care. 
Equation (\ref{Lyapur2}) shows that 
$ \langle v_{\alpha'} v_{\alpha'} \rangle^\circ \sim m^{-1}$, which reflects the equipartition theorem in the 
equilibrium case, while Eq. (\ref{Lyapt1}) suggests that 
$\langle v_\alpha v_\beta \rangle^\circ \sim
\langle x_\alpha v_\beta \rangle^\circ $ for $\alpha \neq \beta$. 
Together with Eq. (\ref{Lyapur1}), this suggests that 
$\langle v_\alpha v_\beta \rangle^\circ \sim \langle x_\alpha v_\beta \rangle^\circ
 \sim m^{0}$ for $\alpha \neq \beta$
and thus that these terms do not necessarily vanish in the mass-less limit $m_\alpha \rightarrow 0$.
This in turn means that  the terms in the boxes in Eqs. (\ref{Lyapt2}) and  (\ref{Lyapur1})
can  be treated perturbatively in an expansion in powers of $m_\alpha$ and
\textcolor{red}{can, to leading order in the particles masses,  be neglected. 
Corrections to the leading-order results for the covariances  can be systematically  
calculated by inserting the leading-order results for the terms in the boxes and by solving the
resulting equations term-by-term in powers of the particle masses. 
Such a calculation of next-leading-order terms would also allow to assess the accuracy of the
leading-order results, but  is rather involved  because for NEQ systems
mixed position-velocity crosscorrelations are present,
as we will show explicitly for the simple case of two coupled particles
in the next section.}
As a main result, we conclude that while the mean-squared velocities of particles 
diverge in the  overdamped limit, the position-velocity correlations between different particles take non-zero 
and finite values for NEQ systems in the overdamped limit.

%%%%%%%%%%%%%%%%%%%%%%%%%%%%%%%%%%%%%%%%%%%%%%%%
\section{Applications}

\subsection{ Two  particles coupled to different temperature reservoirs:
effective temperature concept  and position/momentum localization}
\label{twoparticles}

We now present  explicit results for two particles  that are described by the Newtonian
Hamiltonian as defined generally in Eq. (\ref{Ham2}),
\begin{equation}
 {\cal H}   =  h_{11}x_1^2/2  + h_{22}x_2^2/2  + h_{12}x_1 x_2 + m_1 v_1^2/2 + m_2 v_2^2/2
 \end{equation}
and which are characterized by the  two diagonal  friction coefficients $\gamma_1$ and $\gamma_2$
as defined in Eq. (\ref{Gamma}).
The particles are coupled to two heat reservoirs characterized by inverse thermal energies
$\beta_1$ and $\beta_2$ as defined in Eq. (\ref{LyapBoltz4}).
This is a model system that has been considered by different researchers 
\cite{Filliger2007,Crisanti2012,Dotsenko2013,Berut2016}. We here reproduce  the complete
covariance matrix from our previous calculation\cite{Netz2018}. 
By straightforward solution 
of the set of linear equations (\ref{Lyaps1})-(\ref{Lyapur2})
we obtain to  leading order in the particle masses 
\begin{equation} \label{cov1}
\langle x_1 x_1\rangle^\circ =h_{22} d \left[\frac{1}{\beta_1} + 
\Delta \frac{h_{12}}{ h_{22}} \gamma_2\right] ,
\end{equation}
\begin{equation} \label{cov2}
\langle x_2 x_2 \rangle^\circ =h_{11} d \left[  \frac{1}{\beta_2} - 
\Delta \frac{h_{12}}{ h_{11}} \gamma_1 \right] ,
\end{equation}
\begin{equation}  \label{cov3}
\langle x_1 x_2 \rangle^\circ = - \frac{ h_{12} d}{2} 
\left[ \frac{1}{\beta_1}+ \frac{1}{\beta_2} 
 + \Delta  \frac{   h_{11} \gamma_2 - h_{22} \gamma_1 }{h_{12}} \right],
\end{equation}
\begin{equation}   \label{cov4}
\langle v_1 v_1\rangle^\circ = \frac{1}{m_1}\left[
\frac{1}{ \beta_1} +  \Delta \frac{  h_{12} m_1}{\gamma_1} \right],
\end{equation}
\begin{equation}  \label{cov5}
\langle v_2 v_2\rangle^\circ = \frac{1}{m_2}\left[
\frac{1}{ \beta_2} -  \Delta \frac{  h_{12} m_2}{\gamma_2} \right],
\end{equation}
\begin{equation} \label{cov6}
\langle v_1  v_2\rangle^\circ =  \Delta  \frac{ m_2  h_{11}-m_1 h_{22}}{\gamma_1 m_2 +  \gamma_2 m_1  },
\end{equation}
\begin{equation}\label{cov7}
\langle x_1  v_2\rangle^\circ = -  \langle v_1  x_2\rangle^\circ = \Delta,
\end{equation}
where $d=( h_{11} h_{22} - h^2_{12})^{-1}$ is the inverse determinant of the interaction matrix and
the parameter
\begin{equation}
\Delta = \frac{h_{12}}{ \gamma_1 h_{22} +   \gamma_2 h_{11}}  
 \left( \frac{1}{\beta_2} -  \frac{1}{\beta_1} \right)
\end{equation}
is a measure of the departure from equilibrium.  
For $\Delta =0$, i.e. in equilibrium, the covariance matrix elements are given by the 
inverse of the Hamiltonian matrix according to  Eq. \ref{Ebullet} and in particular 
the off-diagonal velocity coupling terms
$\langle v_1  v_2\rangle^\circ$ and 
the position-velocity correlations
$\langle x_1  v_2\rangle^\circ= - \langle v_1  x_2\rangle^\circ$ vanish.
Off equilibrium, that means for $\Delta \neq 0$, these covariances are non zero
and thus the symmetry of the covariance matrix changes abruptly. 
The expressions in the square brackets 
in Eqs. (\ref{cov1}) -  (\ref{cov5}) 
could in principle  be used to 
define  inverse effective temperatures
for the covariance elements  that are non-zero in equilibrium: inspection of the terms in the square brackets
shows  that  they are all different.
For the covariances Eqs. (\ref{cov6}) and   (\ref{cov7}) 
that are proportional to $\Delta$, the effective inverse  temperatures are also different and diverge 
as equilibrium is
approached, i.e. as $\Delta  \rightarrow 0$. While for one or two of the covariances
an effective temperature can be defined \cite{Grosberg2018},
consideration of the entire covariance matrix shows that the ascription of an effective 
temperature to a particle is not possible. 
This suggests that an effective temperature picture, where one 
assigns effective temperatures to particles, does not describe the particle statistics 
correctly and in particular does not characterize well 
the transition from equilibrium, $\Delta =0$, to NEQ, $\Delta \neq 0$.
To rescue the effective temperature picture one would have to ascribe different 
temperatures to each covariance matrix element, which clearly is far
from the  usefulness of the temperature definition at  equilibrium.
When basing the effective temperature definition on the fluctuation-dissipation relation,
as an  additional effect  a
 frequency dependence  appears \cite{Shraiman1989,Kurchan1997,Puglisi2017,Netz2018},
 which is not reflected in the effective temperatures  one would obtain based on the
 covariance matrix elements. 
 
 As an additional illustration of NEQ effects,
  we calculate the mean-squared difference between
 the particle positions, which from Eqs. (\ref{cov1}) - (\ref{cov3}) is 
 to  order $m^0$ in the particle mass
 given by 
\begin{equation}
\langle (x_1 - x_2)^2  \rangle^\circ =  \frac{ \beta_1 + \beta_2}{2h\beta_1 \beta_2} 
\left[ 1  -  \left( \frac{ \beta_1 - \beta_2}{\beta_1 + \beta_2}  \right)  \left( 
 \frac{ \gamma_1 - \gamma_2}{\gamma_1 + \gamma_2} \right) \right],
\end{equation}
where we have considered two particles whose center of mass is not confined, 
i.e. $h_{11}=h_{22}= -h_{12}=h>0$ (note that this limit must be taken with care since
the interaction matrix $h_{ij}$ is not invertible in this case). 
The first term is the equilibrium result which survives in the limit $\beta_1=\beta_2$. 
Note that 
when the two friction coefficients and the two temperatures are different from each other,
 NEQ effects modify the equilibrium result. In fact, in the limits
$\beta_1/\beta_2  \gg 1 $ and $\gamma_1/\gamma_2  \gg 1 $ 
(or $\beta_2/\beta_1  \gg 1 $ and $\gamma_2/\gamma_1  \gg 1 $)
 the distance between the particles tends to zero, thus indicating positional co-localization of particles,
 which in equilibrium one would only obtain from strong  attractive interactions between particles. 
 Interestingly, since $\beta_\alpha=\gamma_\alpha / b_\alpha^2$, where $b_\alpha$ denotes the
 strength of the Gaussian white noise that enters the Langevin equation, 
 we see that co-localization is automatically obtained when the friction coefficients of 
 particles $\gamma_\alpha$ are modified while keeping the random strengths $b_\alpha$ fixed. 
 This indicates a tendency of particles  to 
 phase separate in position space in NEQ, which is indeed obtained in mixtures of particles
 that are  coupled to different heat baths 
 \cite{Grosberg2015,Frey,Brenner2017,Kremer2017}.
  
 A similar calculation for the mean-squared velocity difference 
 based on Eqs. (\ref{cov4}) - (\ref{cov6}) gives 
\begin{equation} \label{momloc}
\langle (v_1 - v_2)^2  \rangle^\circ =  \frac{ \beta_1 + \beta_2}{m\beta_1 \beta_2} 
\left[ 1  -  \frac{m h^2_{12}}{h \gamma_1 \gamma_2}
 \left( \frac{ \beta_1 - \beta_2}{\beta_1 + \beta_2}  \right)  \left( 
 \frac{ \gamma_1 - \gamma_2}{\gamma_1 + \gamma_2} \right) \right]
\end{equation}
to order $m^0$ and where we used the simplifications $m=m_1=m_2$ and $h_{11}=h_{22}=h$. 
Also here we see that in the limit 
$\beta_1/\beta_2  \gg 1 $ and $\gamma_1/\gamma_2  \gg 1 $ 
(or $\beta_2/\beta_1  \gg 1 $ and $\gamma_2/\gamma_1  \gg 1 $)
the momentum difference between the particles goes down.
 This is indicative of co-localization in momentum space,
meaning that different particles tend to move with the same velocity.
 For a Newtonian Hamiltonian  Eq. (\ref{Ham2}) which is diagonal in momentum space 
 and for which momenta do not couple to  positions, 
 particle velocities are uncorrelated to each other in equilibrium.
We thus conclude that the momentum localization demonstrated in Eq. (\ref{momloc}) is a
NEQ phenomenon  that
has no equilibrium analogue  for Newtonian Hamiltonians.

%%%%%%%%%%%%%%%%%%%%%%%%%%%%%%%%%%%%%%%%%%%%%%%%%
\subsection{Stationary entropy production for three coupled particles:
Heat flux from cold to warm reservoir}
\label{threeparticles}

We will now present an explicit solution for a system consisting  of three coupled particles,
 as schematically visualized in Fig. 2a).
Since here we are only interested in the stationary 
entropy production, Eq. (\ref{Sprod6}), we only need to 
 calculate the stationary velocity-position cross terms  $\langle x_\epsilon v_\alpha \rangle^\circ $.
 We consider the simplified Hamiltonian 
 \begin{equation} \label{HamThree}
 {\cal H} 
 % (x_1, v_1, x_2, v_2, x_3, v_3)   
 =  \frac{h_1}{2} (x_1 - x_2)^2 + \frac{h_3}{2} (x_2 - x_3)^2
 + \frac{m_1}{2} v_1^2    + \frac{m_2}{2} v_2^2     + \frac{m_3}{2} v_3^2
\end{equation}
where only particles 1 and 2 and particles 2 and 3 are coupled via harmonic bonds.
We also assume the friction coefficients to be all the same, i.e. $\gamma_1 = \gamma_2 = \gamma_3 \equiv \gamma$.
The solution strategy consists in  eliminating $\langle v_{\alpha'}^2 \rangle^\circ$ 
by inserting Eq. (\ref{Lyapur2}) into Eq.  (\ref{Lyapt2}), which results in 3 equations,
and solving the six equations defined by Eq. (\ref{Lyapur1}) by using Eqs. (\ref{Lyaps1}) and (\ref{Lyaps2}).
These are nine equations for nine unknowns, it turns out that Eq. (\ref{Lyapt1}) is not needed.
The results are  to  order $m^0$ in the particle mass given by 
\begin{equation} \label{x2v3}
\langle x_2 v_3 \rangle^\circ = \frac{1}{6 \gamma (h_1 + h_3)}
\left[ \frac{h_1}{\beta_1} +  \frac{h_1+ 3h_3 }{\beta_2}-\frac{2 h_1+ 3h_3 }{\beta_3} \right],
\end{equation}
\begin{equation}  \label{x1v2}
\langle x_1 v_2 \rangle^\circ = \frac{1}{6 \gamma (h_1 + h_3)}
\left[    \frac{2 h_3+ 3h_1 }{\beta_1} -      \frac{h_3}{\beta_3} -   \frac{h_3+ 3h_1 }{\beta_2}     \right],
\end{equation}
\begin{equation}
\langle x_1 v_3 \rangle^\circ = \frac{1}{ \gamma \Theta}
\left[    \frac{h_3+ h_2 }{\beta_1} +      \frac{h_1 - h_3}{\beta_2} -   \frac{h_1+ h_2 }{\beta_3}     \right],
\end{equation}
where we defined for convenience
\begin{equation}
 \Theta \equiv 
 h_1 + 2 h_2 + h_3 +\frac{h_2^2- h_3^2}{h_1}+\frac{h_2^2- h_1^2}{h_3}.
\end{equation}
Inserting these results into the expression for the entropy production Eq. (\ref{Sprod6}) we obtain
\begin{eqnarray}
	\label{Sprod7}
\dot{\cal S}_{\rm res}^\circ / k_B  = 
\frac{1}{2}
h_{1}\langle x_2 v_1 \rangle^\circ (\beta_1 - \beta_2 )
+\frac{1}{2}
 h_{3}\langle x_2 v_3 \rangle^\circ (  \beta_3 - \beta_2 ) \nonumber   \\
=\frac{1}{12 \gamma (h_1+h_3)} \left[
 h_1  h_3\left(\sqrt{\frac{\beta_1}{\beta_3}}-  \sqrt{\frac{\beta_3}{\beta_1}} \right)^2    \right.   \hspace{1cm}   \\
 +  h_1(3h_1+h_3)\left(\sqrt{\frac{\beta_1}{\beta_2}}-  \sqrt{\frac{\beta_2}{\beta_1}} \right)^2  \hspace{1cm}  \nonumber \\
 \left. + h_3(3h_3+h_1)\left(\sqrt{\frac{\beta_3}{\beta_2}}-  \sqrt{\frac{\beta_2}{\beta_3}} \right)^2 \hspace{1cm} \nonumber
\right].
\end{eqnarray}
Obviously, the  entropy production is positive and  finite in the zero mass limit if the reservoir 
temperatures are different and if the coupling strengths $h_1$ and $h_3$ are finite
and positive.

The  limiting case of  $h_3=0 $  is insightful: in this case, particle 3  becomes decoupled
and we are basically left with only two coupled particles.
The expression Eq. (\ref{Sprod7}) simplifies to 
\begin{equation}
	\label{Sprod8}
\dot{\cal S}_{\rm res}^\circ / k_B  =   \frac{h_1 }{4 \gamma}  
\left(\sqrt{\frac{\beta_1}{\beta_2}}-  \sqrt{\frac{\beta_2}{\beta_1}} \right)^2 
\end{equation}
which describes the stationary entropy production of two reservoirs of inverse temperatures 
$\beta_1$ and 
$\beta_2$ that act on two particles that are subject to friction with  coefficients $\gamma$ 
and which are coupled by a  harmonic spring of strength $h_1$. 
Obviously, also this entropy production is never negative and finite if the reservoir 
temperatures are different and if the coupling strength $h_1$ is finite.

The case of three particles that are coupled to heat reservoirs at three different temperatures 
allows to demonstrate
as an interesting NEQ effect  the pumping of heat against a temperature gradient. Before we go into
the analysis, we note that this   
does not constitute a violation of the second law of thermodynamics but  rather is a NEQ
entrainment  effect. 
To proceed, we calculate the stationary heat flux from the heat reservoir at temperature $T_1$ to particle 1,
which according to Eqs. (\ref{Sprod2}) and  (\ref{x1v2}) is given by
\begin{eqnarray}
	\label{Sprod9} 
\dot{Q}_{1}^{\circ}    = &
-  h_{1}\langle x_2 v_1 \rangle^\circ =   h_{1}\langle x_1 v_2 \rangle^\circ \nonumber   \\
= & \frac{h_1}{6 \gamma (h_1 + h_3)}
\left[    \frac{2 h_3+ 3h_1 }{\beta_1} -      \frac{h_3}{\beta_3} -   \frac{h_3+ 3h_1 }{\beta_2}     \right].
\end{eqnarray}
The stationary heat flux from the heat reservoir at temperature $T_3$ to particle 3 is
 according to Eqs. (\ref{Sprod2}) and  (\ref{x2v3})  given by
\begin{eqnarray}
	\label{Sprod11} 
\dot{Q}_{3}^{\circ}    = &
 - h_3   \langle x_2 v_3 \rangle^\circ  \nonumber   \\
=&  - \frac{h_3}{6 \gamma (h_1 + h_3)}
\left[    \frac{h_1}{\beta_1} +      \frac{h_1+3h_3}{\beta_2}  -  \frac{2h_1+ 3h_3 }{\beta_3}     \right].
\end{eqnarray}
Due to energy conservation $\dot{Q}_{2}^{\circ} = -  \dot{Q}_{1}^{\circ} -\dot{Q}_{3}^{\circ} $ holds.
Note that all stationary heat fluxes  $\dot{Q}_{1}^{\circ}$,  $\dot{Q}_{2}^{\circ}$, $\dot{Q}_{3}^{\circ}$ 
obviously vanish in equilibrium, i.e. when the temperatures of all heat reservoirs  are the same.
The  flux from heat reservoir 1 according to Eq. (\ref{Sprod9}) is negative, $\dot{Q}_{1}^{\circ} <0$,
i.e. heat flows into  reservoir 1,  for 
\begin{equation} \label{pump1}
\frac{T_3}{T_2} > - 1 - 3h_1/h_3 +
(2+3h_1/h_3)\frac{T_1}{T_2},
\end{equation}
which for equal elastic coupling strengths $h_1=h_3$ simplifies to 
\begin{equation} \label{pump2}
\frac{T_3}{T_2} > - 4  + 5 \frac{T_1}{T_2}.
\end{equation}
The  flux from heat reservoir 3 according to Eq. (\ref{Sprod11}) is positive, $\dot{Q}_{3}^{\circ} > 0$,
i.e. heat flows out of   reservoir 3, for 
\begin{equation} \label{pump3}
\frac{T_3}{T_2} > \frac{ h_1+ 3 h_3}{2h_1 + 3 h_3}
+ \frac{h_1}{2 h_1 + 3 h_3 } \frac{T_1}{T_2},
\end{equation}
which for equal elastic coupling strengths $h_1=h_3$ simplifies to 
\begin{equation} \label{pump4}
\frac{T_3}{T_2} > 4/5  + \frac{T_1}{5 T_2}.
\end{equation}
As expected, the conditions Eqs. (\ref{pump1}) and (\ref{pump3}) are  
simultaneously satisfied, i.e. heat flows into reservoir 1 and  out of  reservoir 3, for 
\begin{equation} \label{pump5}
T_1 <   T_2  < T_3,
\end{equation}
 i.e.,  per unit time  the  heat amount  $|\dot{Q}_{3}^{\circ}|$,
 flows from particle 3 (which is coupled to the hot heat reservoir) to particle 2 (which is coupled
 to the warm reservoir), and the heat amount $|\dot{Q}_{1}^{\circ}|$
 flows from particle 2 to particle 1 (which is coupled to the cold reservoir). 
 
 More interestingly,  the conditions Eqs. (\ref{pump1}) and (\ref{pump3}) can also be   
simultaneously satisfied for the scenario
\begin{equation} \label{pump6}
T_2 <   T_1  < T_3.
\end{equation}
 In this case there is a small range of temperatures 
where heat flows from particle 3 (coupled to the hot heat reservoir) to particle 2 (coupled
 to the cold reservoir), which is expected, but at the same time 
 a heat amount $|\dot{Q}_{1}^{\circ}|$
 flows from particle 2  (coupled
 to the cold reservoir) to particle 1 (coupled to the warm reservoir). 
A similar scenario is provided for 
\begin{equation} \label{pump6b}
T_1 <   T_3  < T_2,
\end{equation}
where for a small range of temperatures heat flows from particle  2 (coupled to the hot heat reservoir) to particle 1 (coupled
 to the cold reservoir), which is expected, but at the same time 
 a heat amount $|\dot{Q}_{3}^{\circ}|$
 flows from particle 3  (coupled
 to the warm  reservoir) to particle 2 (coupled to the hot reservoir). 
 This means we find situations where heat flows against the temperature gradient of the heat reservoirs 
 that are coupled to the particles. 
This situation is indicated in Fig. 2b)
in a  phase diagram for equal elastic coupling $h_1=h_3$, in which case the 
simplified inequalities  Eqs. (\ref{pump2}) and (\ref{pump4}) hold.
In the phase diagram the inequalities Eqs. (\ref{pump2}) and (\ref{pump4}) are indicated by straight lines,
and the region where  Eq.  (\ref{pump5}) holds is indicated in green.
The regions where the inequalities Eqs. (\ref{pump2}) and (\ref{pump4}) 
and in addition    Eq.  (\ref{pump6}) or  Eq.  (\ref{pump6b}) hold are indicated in red.
Our discussion uses the fact that any heat  $\dot{Q}_{1}^{\circ}$ that is transferred to or from reservoir 1 
is transferred between particles 1 and 2 via elastic interactions, likewise,
any heat  $\dot{Q}_{3}^{\circ}$ that is transferred to or from reservoir 3
is transferred between particles 2 and 3.
In other words, we do not only know the stationary heat fluxes from the reservoirs but also the stationary energy  fluxes
between the particles, which is due to the simple linear topology of the elastic particle interactions as indicated in Fig. 2a).

As mentioned before, the finding of a heat flux against the reservoir temperature gradient
does not violate the second 
 law of thermodynamics.
 Firstly, the heat flux from the particle coupled to the cold heat bath to 
the particle coupled to the warm heat bath is accompanied by an even larger heat flux from the 
particle coupled to the hot heat bath to the particle coupled to the warm heat bath, 
our result for the total entropy production Eq. (\ref{Sprod7}) is strictly positive. 
Secondly, heat is not transferred directly between the heat reservoirs but only between
particles that are coupled to heat reservoirs. In this connection it is crucial 
to  remember (according to our previous
discussion centered around the covariance elements Eqs. (\ref{cov1}) - (\ref{cov7}))
 that  particles are not characterized by the temperature
 of the heat reservoir they are coupled to. 
 Therefore, 
 since the particles  do not have well-defined
  effective temperatures, the second law of thermodynamics is not  violated. 
  One could be tempted to define effective temperatures based on the heat fluxes,
  but such a definition would be based solely on  one entry of the covariance matrix,
  namely the off-diagonal position velocity coupling $\langle x_\alpha v_\beta \rangle^\circ$,
  and not work for other applications.

\begin{figure*}	
	\centering
	\includegraphics[width=12cm]{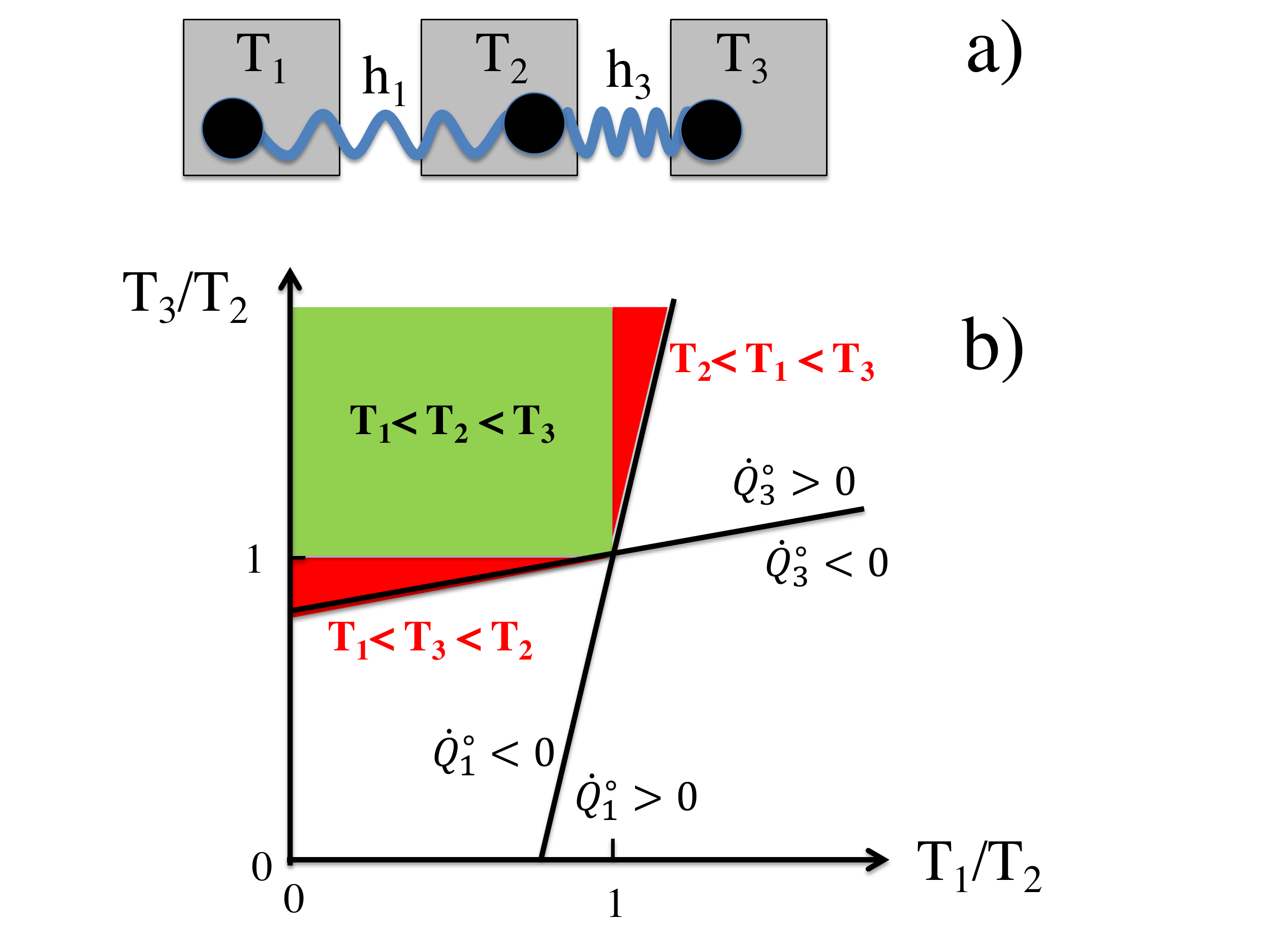}
\caption{ 
a) Schematic representation of three particles  that interact  elastically  
and  that are coupled  to heat baths with different temperatures.
b) Stationary heat flux  diagram for symmetric elastic couplings $h_1=h_3$,
showing a  colored region where heat   flows from particle 3 to particle 2,
i.e.  $\dot{Q}_{3}^{\circ} > 0$, and heat flows from particle 2 to particle 1, i.e.   $\dot{Q}_{1}^{\circ} < 0$.
The green area denotes the  temperature range   $T_1 <  T_2 <T_3$ 
and the red areas denote 
$T_2 <  T_1 <T_3$  and  $T_1 <  T_3 <T_2$.
The green area is the expected temperature range where heat flows from  particle 3 coupled to 
the hot reservoir at $T_3$ to particle 2 coupled to the warm reservoir at $T_2$
and from particle 2 coupled to the warm reservoir at $T_2$ to particle 1 coupled to the cold
reservoir at $T_1$.
In the red areas heat flows between the particles against the temperature gradients of the heat reservoirs:
 In the upper red area heat flows from  particle 2 coupled to the cold
reservoir at $T_2$ to  particle 1 coupled to the warm  reservoir at $T_1$,
 in the lower red area heat flows from  particle 3 coupled to the warm
reservoir at $T_3$ to  particle 2 coupled to the hot  reservoir at $T_2$.
		}
	\label{fig2}
\end{figure*}

\subsection{Mapping on active particles} \label{activeparticles}

Similar to recent calculations\cite{Ramaswamy2018}, we here  show how 
active particle models can be described within the current framework of 
Markovian coupled particles.
We want to describe a single active particle, for this 
we reduce the Hamiltonian Eq. (\ref{HamThree})
to  two coupled massive particles and obtain 
 \begin{equation}
 {\cal H} 
 % (x_1, v_1, x_2, v_2, x_3, v_3)   
 =  \frac{h_1}{2} (x_1 - x_2)^2 
 + \frac{m_1}{2} v_1^2    + \frac{m_2}{2} v_2^2.
\end{equation}
By  assuming the friction coefficients to be the same, i.e.
 $\gamma_1 = \gamma_2  \equiv \gamma$,
 and choosing two different heat bath temperatures $\beta_1$ and $\beta_2$, 
we obtain from Eqs.  (\ref{Langevin2}),  (\ref{A2}),   (\ref{Ham2}),  (\ref{Gamma})
  the coupled set of linear Langevin equations 
\begin{eqnarray}
	\label{Lang}
m_1\, \dot{v_1}(t) &=& - \gamma v_1(t)  - h_1 [ x_1(t)-x_2(t)]  + \sqrt{ \gamma /\beta_1} F_1(t),  \nonumber \\
m_2 \, \dot{v}_2(t) &=& - \gamma v_2 (t) - h_1 [ x_2(t)-x_1(t)]  + \sqrt{ \gamma /\beta_2} F_2(t).  \nonumber
\end{eqnarray}
The Langevin equation for the second particle is straightforwardly  solved in the mass-less limit
$m_2=0$ and gives
\begin{equation}
x_2(t) = \int_{-\infty}^t {\rm d}t' {\rm e}^{-(t-t')h_1/\gamma}\left[ \frac{h_1}{\gamma} x_1(t') + 
(\beta_2 \gamma)^{-1/2}  F_2(t') \right].
\end{equation}
Inserting this solution into the Langevin equation  for the first particle,
we obtain  the generalized Langevin equation 
\begin{equation} \label{gle}
m_1 \dot{v_1} (t) = -
\int_{-\infty}^\infty {\rm d}t' \Gamma(t-t') v_1(t') +F_R(t).
\end{equation}
The memory function that appears in  Eq. (\ref{gle})  is given by 
\begin{equation} \label{memory}
\Gamma (t) = \theta(t)  \left[ 2 \gamma \delta(t) +
h_1   {\rm e}^{- t h_1/\gamma}\right] 
\end{equation}
where $\theta(t)$ denotes the Theta function with the properties
$\theta(t)=1$ for $t>0$ and $\theta(t)=0$ for $t<0$, which makes 
the memory function  single-sided. 
The noise $F_R(t)$   in  Eq. (\ref{gle})  is given by 
\begin{equation}
F_R (t) =   \sqrt{ \gamma /\beta_1}  F_1(t) + \frac{h_1}{ \sqrt{ \beta_2 \gamma}}
 \int_{-\infty}^t {\rm d}t' 
 {\rm e}^{- (t-t')  h_1/\gamma} F_2(t')
\end{equation}
and consists of the noise acting directly on the first  particle 
and a term due to noise acting on the second particle.
 The latter term consists of a  convolution integral because this noise
 is  transmitted  via the elastic bond  of strength $h_1$.
 Defining the auto-correlation function of the random noise as 
 $C_{\rm FF}(t)=\langle F_R(0) F_R(t) \rangle$, we obtain \cite{Netz2018}
\begin{equation} \label{CFF}
\beta_1 C_{\rm FF}(t)   =   
 2 \gamma \delta(t) +
h_1(\beta_1/\beta_2) {\rm e}^{-|t|h_1/\gamma}.
\end{equation}
 Comparing Eq. (\ref{memory}) and Eq. (\ref{CFF}), we see that 
 $\beta_1 C_{\rm FF}(t)  = \Gamma (|t|)$, a consequence of the standard 
 fluctuation-dissipation theorem \cite{Risken}, only holds for 
 $\beta_1 = \beta_2$, i.e., if the two reservoir temperatures are the same. 
 If $\beta_1 \neq \beta_2$,
 the memory kernel $\Gamma(t)$  and the random force correlation function  $C_{\rm FF}(t)$
 differ, which points to FDT violation and thus to the presence of an active NEQ process.
 The equation of motion Eq. (\ref{gle}) is similar to previously studied active particle models 
 that were shown to exhibit NEQ phase transitions \cite{Marchetti2012,Farage2015}.
 In fact, the active Ornstein-Uhlenbeck model is obtained by setting the exponential
 in the memory function Eq. (\ref{memory})  to zero while keeping the  exponential 
 term  in the noise correlator Eq. (\ref{CFF}). 
 The entropy production of such active particle models has been intensely studied and
 debated \cite{Fodor2016,Mandal2017,Marchetti2018,Dabelow2019}.
The entropy production of the present active particle model,
defined by Eq. (\ref{gle}) in conjunction with Eqs. (\ref{memory}) and (\ref{CFF}),
is given exactly by the simple expression
Eq. 	(\ref{Sprod8}); alternatively, it can  be obtained directly from simulations by evaluating
the position-velocity correlation functions in the  general expression Eq. (\ref{Sprod6}).
The advantage of the present active particle model, which  follows from a Newtonian Hamiltonian,
is that  the extremal and stability conditions 
in terms of the free entropy functional 
derived in this work are valid and not only describe the stationary NEQ distribution
itself but also the approach to the stationary NEQ distribution.
We hasten to add that not all active particle models can be mapped on our model, in particular
models with non-Gaussian velocity distributions are not captured by our linear equations.
In future work interacting active particles similar to the model derived here will be studied 
analytically as well as in simulations, it will be interesting to see whether the NEQ position
and momentum localization effects demonstrated in Sect. \ref{twoparticles}
also show up in those more complex systems.

\section{Conclusions}

In this paper we consider the approach of Hamiltonian many-body systems 
that are coupled to multiple heat reservoirs with different temperatures
to stationary NEQ distributions.
Based on the exactly calculated approach of the covariance matrix  $E(t)$ to 
its stationary NEQ form  $E^\circ$, we construct the functional 
${\cal S}_{\rm free} (t)$ that yields the stationary covariance matrix  $E^\circ$
at its extremum with respect to variations of $E(t)$. This function
is called  the free entropy, and  it  consists of the system distribution entropy ${\cal S}(t)$
and a term that accounts for interactions within the system, described
by the Hamiltonian, and the frictional and noise coupling to the heat reservoirs. 

Since in the stationary state the free entropy production  by construction vanishes, 
as explained  in Section \ref{extremalprop},
the difference between the free entropy production and the total entropy production
is the reservoir entropy production in the stationary state, $ \dot{\cal S}^\circ_{\rm res}$,
which is constant in time. 
 The total entropy production $\dot {\cal S}_{\rm tot}(t)$, which is the sum of the system entropy production 
$\dot {\cal S}(t)$
and the reservoir entropy production $ \dot {\cal S}_{\rm res}(t)$, follows from Eq. (\ref{intro1}) by
differentiation as
\begin{equation}
	\label{STOT2}
\dot {\cal S}_{\rm tot}(t) = \dot {\cal S}(t)+ \dot {\cal S}_{\rm res}(t)=
\dot  {\cal S}_{\rm free} (t)+  \dot{\cal S}^\circ_{\rm res}.
  \end{equation}

Using the result in Eq.  (\ref{Sfree4}) we obtain the explicit expression
\begin{equation}
\label{STOT4}
\frac{ \dot{\cal S}_{\rm tot}(t)}{k_B}   =  \frac{\dot{\cal S}(t)}{k_B} -  \frac{ E^{\circ -1}_{ij}  \dot E_{ij}(t)  }{2}  
-   E^{\circ -1}_{ij}    \langle \dot z_i(t)  \rangle  \langle   z_j(t)\rangle +  \dot{\cal S}^\circ_{\rm res},
\end{equation}
which consists of the system distribution entropy production $\dot{\cal S}(t)$,
the stationary reservoir entropy production $\dot{\cal S}^\circ_{\rm res}$, 
and two   terms  that account for interactions within the system
as well as  the frictional and noise coupling to the environment. 
NEQ effects make these coupling terms differ dramatically from their equilibrium 
counterparts, in which case  $E^{\circ -1}_{ij}$ becomes replaced by the Hamiltonian matrix $H_{ij}/(k_BT)$.
In Appendix \ref{AppE} we attempt to derive 	Eq. (\ref{STOT2}) explicitly
by calculating  the  time-dependent reservoir entropy production $\dot {\cal S}_{\rm res}(t)$
from our explicit expressions for the time-dependent heat fluxes between the heat  reservoirs and 
the system. We demonstrate that the heat fluxes are by themselves not sufficient  
to derive the reservoir entropy production $\dot {\cal S}_{\rm res}(t)$ if the reservoir temperatures are not the same,
which implies that internal reservoir degrees of freedom contribute in a nonneglible 
manner to the reservoir entropy production for a NEQ system.
The connection between the free entropy  production  $\dot  {\cal S}_{\rm free} (t)$ (excluding the reservoirs) 
and the total entropy production $\dot {\cal S}_{\rm tot}(t) $  (including the reservoirs) will be reconsidered  in future work 
using microscopic models for the heat reservoirs.

It is important to note that the reservoir entropy production  in the stationary state,
denoted as $\dot{\cal S}^\circ_{\rm res}$ 
and given explicitly in   Eq. (\ref{Sprod6}),
does not depend on the time-dependent covariance matrix  $E(t)$
but only on the stationary covariance matrix  $E^\circ$, so it is a constant
with respect to variations in $E(t)$; thus,  the total entropy production $\dot{\cal S}_{\rm tot}(t)$
exhibits the identical extremal and stability properties as the free entropy 
production $\dot{\cal S}_{\rm free}(t)$. 
It trivially follows from 
Eq. (\ref{STOT4}) that the total entropy ${\cal S}_{\rm tot}(t) $ of a NEQ system increases indefinitely with time,
as shown explicitly in Eq. (\ref{intro1}),
and thus is not bounded, whereas the free entropy $ {\cal S}_{\rm free} (t)$  is a well-defined and finite 
expression even for NEQ systems. 
Expression Eq. (\ref{STOT4}) 
will be useful for various applications whenever the total entropy production of 
different NEQ states of a system need to be compared.
The functional Eq. (\ref{STOT4}) should also
allow to construct approximate methods for the description of 
NEQ  non-linear systems as well as for NEQ phase transitions. 

One  advantage of the current  formulation is that in the limit when  all
heat reservoir temperatures become equal and the  NEQ system transforms into 
an equilibrium system, the NEQ free entropy smoothly crosses over to the 
equilibrium free energy divided by $-T$, which displays the standard
equilibrium extremal and stability properties of a canonical system.
We thus have derived a unified framework to treat NEQ systems that are coupled 
to heat reservoirs at different temperatures on the same footing as 
equilibrium systems.
We have in our work restricted ourselves to one specific class of NEQ models,
namely where NEQ is produced by stochastic forces with vanishing mean. 
Forces with a non-vanishing mean are straightforward to include and will 
be treated in future work.

Our approach rests on the harmonic approximation for the 
interaction Hamiltonian and for the friction and stochastic terms. 
It is not clear how to extend the present derivation to non-linear systems,
here variational and perturbative methods will most likely be helpful.
 Clearly, a harmonic model can always be obtained from a more complex, non-linear
 model  by a saddle-point  expansion in terms of suitably defined deviatory coordinates.
 Our model should thus also apply to non-linear  systems as long as
 this saddle-point approximation is justified.

\begin{acknowledgments}
We acknowledge funding from the  DFG  via the SFB 1114,
from the European Research Council (ERC) under the European Union’s Horizon 2020 research and innovation program 
(grant agreement No. [835117]) and from the Infosys Foundation.

\end{acknowledgments}

\begin{appendix}

\section{Derivation of Shannon entropy} \label{AppA}

We start from the canonical partition function 
\begin{equation} \label{partition}
{\cal Z}= \int {\rm d} \vec{z}  {\rm e}^{ -  \beta {\cal H} (\vec{z})}
\end{equation}
where $\beta = 1/(k_BT)$ denotes the inverse thermal energy.
 Using the thermodynamic definition of the free energy ${\cal F}  =- k_BT  \ln {\cal Z} $ 
 we can write  
 \begin{equation} \label{partition2}
{\cal F}= -  k_BT  \ln   \int {\rm d} \vec{z}  {\rm e}^{ -  \beta {\cal H} (\vec{z})}.
\end{equation}
From this we obtain, using the 
thermodynamic definition ${\cal S}  = - \partial {\cal F}  /\partial T$,
 for the entropy
 \begin{equation} \label{partition3}
{\cal S}= 
 k_B   \ln  {\cal Z} +
 {\cal Z}^{-1} T^{-1} 
 \int {\rm d} \vec{z}  {\cal H} (\vec{z})  {\rm e}^{ -  \beta {\cal H} (\vec{z})}.
\end{equation}
From the definition of the  normalized equilibrium  distribution 
 Eq.(\ref{Boltz}), we obtain 
\begin{equation} \label{partition4}
 \beta {\cal H} (\vec{z}) = - \ln \rho(\vec{z}) - \ln {\cal Z},
 \end{equation}
which is inserted into Eq.(\ref{partition3}) to give
 \begin{equation} \label{partition3b}
{\cal S}= 
- k_B  {\cal Z}^{-1}  
 \int {\rm d} \vec{z}   {\rm e}^{ -  \beta {\cal H} (\vec{z})} \ln \rho(\vec{z}).
\end{equation}
Using again the definition of the  normalized equilibrium  distribution 
 Eq.(\ref{Boltz}), we finally obtain 
 the Shannon expression for the entropy  as
\begin{equation}
{\cal S}  /k_B   = -   \int {\rm d} \vec{z}  \rho(\vec{z})  \ln \rho(\vec{z}).
\end{equation}

\section{The time dependent probability distribution is Gaussian}  \label{AppB}

Here  we show that the time-dependent distribution function
is Gaussian and governed by the time-dependent  inverse covariance matrix.
The calculation holds for equilibrium as well as for NEQ  systems.
Given a solution  $\vec{z}(t)$ of the Langevin Eq. (\ref{Langevin}),
we construct the time dependent probability distribution as
 \begin{equation} \label{Gauss1}
\rho( \vec{z'},t) =    \vec{\delta}( \vec{z'} -  \vec{z}(t)).
\end{equation}
Using the Fourier representation of the delta function and the time-dependent solution
Eq. (\ref{sol1}) for given initial value,   we obtain 
\begin{widetext}
  \begin{equation} \label{Gauss2}
\rho_F[ \vec{z'},t,\vec{F}(\cdot)] =  
\int \frac{{\rm d} \vec{\omega}}{(2\pi)^{2N}} 
\exp \left\{
- \imath \omega_i z_i'  + \imath \omega_i   \left[  \langle z_i (t) \rangle
+   \int_0^t {\rm d} t'   {\rm e}^{ - (t-t') A }_{ij}  \Phi_{jk} F_k  (t') \right]\right\},
\end{equation}
which is a functional of the  random force trajectory $\vec{F}(t)$.
The probability distribution is obtained by a path integral over all random force trajectories
as 
 \begin{equation} \label{Gauss3}
 \rho( \vec{z'},t) =
\int {\cal D}\vec{F}(\cdot) P[\vec{F}(\cdot)]
\rho_F[ \vec{z'},t,\vec{F}(\cdot)].
\end{equation}
Here $P(\vec{F}(\cdot))$ is the path integral weight, which for 
diagonal white noise is given by
 \begin{equation} \label{Gauss4}
P[\vec{F}(\cdot))] = {\cal N}_F^{-1} 
\exp \left\{ -\frac{1}{4} \int {\rm d} t  {\rm d} t' F_k(t) \delta(t-t') F_k(t')   \right\}
\end{equation}
and leads  to 
$\langle F_i  (t) \rangle = 0$ and   
$\langle F_i  (t) F_j (t')\rangle = 2\delta_{ij} \delta(t-t')$, where
the normalization factor is given by ${\cal N}_F$.
The path integral over the random noise in Eq.(\ref{Gauss3}) can be performed
and leads to the expression 
  \begin{equation} \label{Gauss5}
\rho( \vec{z'},t) =  
\int \frac{{\rm d} \vec{\omega}}{(2\pi)^{2N}} 
\exp \left\{
- \imath \omega_i z_i'  + \imath \omega_i    \langle z_i (t) \rangle
-\omega_i E_{ij}(t) \omega_j   \right\},
\end{equation}
where the covariance matrix $E_{ij}(t)$  is given by Eq.(\ref{sol3}).
Performing the integral over $\vec{\omega}$ leads to 
\begin{equation} \label{rhoapp}
\rho( \vec{z'},  t )  =  
{\cal N}^{-1} (t)  \exp  \left\{-    [z'_i  - \langle z_i (t) \rangle ]       
E^{-1}_{ij}(t)  [z'_j  - \langle z_j (t) \rangle]   /2\right\},
\end{equation}
with ${\cal N}$ given by Eq. (\ref{N}), which is identical to the expression given in Eq. (\ref{rho}).
Since the mean state vector $\langle z_i (t) \rangle$ in Eq. (\ref{rhoapp}) depends according to Eq. (\ref{sol2}) 
on the initial state vector  $z_i (0)$, the time-dependent probability distribution we derived here is in fact the
conditional distribution and thus corresponds to the 
 Green's function, this can be made explicit by rewriting Eq. (\ref{rhoapp}) as
\begin{equation} \label{Green}
\rho( \vec{z'},  t | \vec{z},  0)  =  
{\cal N}^{-1} (t)  \exp  \left\{-    [z'_i  -  {\rm e}^{ - t A }_{ik} z_k ]       
E^{-1}_{ij}(t)  [z'_j  -  {\rm e}^{ - t A }_{jl} z_l ]   /2\right\}.
\end{equation}
\end{widetext}

\section{Semi-positive definiteness of matrix product trace}  \label{AppC}

We  start from the part of the expression Eq. (\ref{freeenderiv3}) 
for the free energy production rate of an equilibrium system that involves 
the trace of a product of four matrices, 
\begin{equation}
\label{posdef1}
{\cal \dot{F}}  =  
-  C_{km} [ H_{ml} - k_BT  E^{-1}_{ml} ]\frac{E_{lj}}{k_BT}    [ H_{jk} - k_BT  E^{-1}_{jk} ],
\end{equation}
where we have for simplicity omitted all time dependencies. 
An analogous  expression also appears in the free entropy production
in Eq. (\ref{Sfree5}). The matrix trace expressions that appear in the 
free energy Eq. (\ref{freeen2b}) and the free entropy Eq.  (\ref{Ssecond})
are special cases of the more general matrix product trace we consider here.

By defining the matrix $M_{ml} = H_{ml} - k_BT  E^{-1}_{ml}$, the
expression can be written more compactly as 
\begin{equation}
\label{posdef2}
k_BT  {\cal \dot{F}}  =  
-  C_{km} M_{ml}  E_{lj}    M_{jk} 
\end{equation}
where all matrices $C, M, E$ are symmetric and assumed to be non-defective,
 $E$ is positive definite, 
$C$ is semi-positive definite and the definiteness of $M$  (since it is 
the difference of two matrices) is not specified. 
We first diagonalize the matrix $E$ by the similarity transformation
\begin{equation}
\label{posdef3}
E P^E P^{E-1} = P^E D^E P^{ E-1}=  P^E D^E P^{E,T} 
\end{equation}
where in the last step we used that $E$ is symmetric and $P^E$ is thus 
an orthogonal matrix.  The matrix $D^E_{i'j'}= \delta_{i'j}d^E_{i'}$
is diagonal with diagonal elements  $d^E_{i'}$.
We thus obtain
\begin{equation}
\label{posdef4}
k_BT  {\cal \dot{F}}  =  
-  C_{km} M_{ml}P_{li}D^E_{ij}P_{nj}M_{nk}.
\end{equation}
We next define $N_{mi}=M_{ml}P_{li}$
and obtain, by using that $M$ is symmetric,  
\begin{equation}
\label{posdef5}
k_BT  {\cal \dot{F}}  =  
-C_{km}N_{mi}D^E_{ij}N_{kj}.
\end{equation}
We next diagonalize the matrix $C$ by the similarity transformation
\begin{equation}
\label{posdef6}
C P^C P^{C-1} = P^C D^C P^{ C-1}=  P^C D^C P^{C,T},
\end{equation}
where $D^C_{i'j'}= \delta_{i'j}d^C_{i'}$
and obtain
\begin{equation}
\label{posdef7}
k_BT  {\cal \dot{F}}  =  
-P^C_{kl}D^C_{lo}P^C_{mo}N_{mi}D^E_{ij}N_{kj}.
\end{equation}
We now  define $R_{oi}=P^C_{mo} N_{mi}$, which is equivalent to 
$R_{lj}=  N_{kj}  P^C_{kl}$,
and thus obtain 
\begin{equation}
\label{posdef8}
k_BT  {\cal \dot{F}}  =  
-D^C_{lo}R_{oi}D^E_{ij}R_{lj}.
\end{equation}
Now using that $D^C$ and $D^E$ are diagonal matrices we obtain
\begin{equation}
\label{posdef9}
k_BT  {\cal \dot{F}}  =  
-  d^C_l R_{li} d^E_i R_{li}= 
-d^C_l d^E_i R^2_{li} \leq 0,
\end{equation}
where the inequality follows since the matrix elements of $R$ are real 
and all eigenvalues of the matrices $C$ and $E$ are not negative.

\section{Condition of detailed balance}  \label{AppEnew}
Detailed balance is satisfied if the probability for a transition from a state vector $ \vec{z}$ at time 
$t$ to a state vector $ \vec{z}'$ at time  $t+ \tau $ is the same as the probability for a transition from 
 $ \vec{z}'$ at time  $t $ to  $ \vec{z}$ at time  $t+ \tau $, provided that all velocities are reversed. Detailed balance is satisfied 
 for systems that are in equilibrium and is   standardly used as the definition of equilibrium \cite{deGroot}.
  In this section we will
 demonstrate that the condition of detailed balance is equivalent to the condition based on the Boltzmann distribution,
 provided that the friction matrix is symmetric and that the stochastic field correlations satisfy certain 
 symmetry relations. Our calculation is similar to classical derivations
 \cite{Kampen1957a,Kampen1957b,Graham1971,Schnakenberg1976}.
 
Using the  joint probability distribution, the detailed balance condition can be written as  \cite{Kampen1957a,Kampen1957b,Graham1971}
\begin{equation} \label{det1}
\rho( \vec{z}',  t ; \vec{z},  0)  =  \rho( W \vec{z} ,  t ; W  \vec{z}',  0)
\end{equation}
where $W$ is the diagonal matrix that reverses all velocities and is given by 
 \begin{equation} \label{W}
W = \begin{pmatrix}
 1 & 0  &0  &0 & \\
0  & -1 & 0 & 0 & \\
 0 & 0 &1  &0 & \\
0 & 0 & 0 & -1 & \\
    &    &    &    & \ddots 
\end{pmatrix}	.
\end{equation}
The detailed-balance condition can be brought into its standard form by using the conditional probability, resulting in 
\begin{equation} \label{det2}
\rho( \vec{z}',  t | \vec{z},  0)  \rho^\circ( \vec{z}) =  \rho( W \vec{z} ,  t | W  \vec{z}',  0) \rho^\circ( \vec{z} '),
\end{equation}
where 
\begin{equation} \label{det2b}
 \rho^\circ( \vec{z}) =  \rho( \vec{z},t \rightarrow \infty) = 
  {\cal N}^{\circ -1}   \exp  \left\{-   z_i      
E^{\circ -1}_{ij}   z_j   /2\right\}
 \end{equation}
 is the stationary distribution and accordingly the time argument is omitted. 
 The stationary normalization constant follows from Eq. (\ref{N})  as 
$  {\cal N}^\circ = \sqrt{(2\pi)^{2N} \det E^\circ}$.
The conditional probability distribution $\rho( \vec{z}',  t | \vec{z},  0)$, which is equivalent to the Green's function of the
stochastic problem defined by the Langevin or the Fokker-Planck equation, is explicitly given in Eq. (\ref{Green}) and is valid 
for equilibrium as well as for non-equilibrium systems.
With the  definition
\begin{equation} \label{det3}
 \vec{z}' =  \vec{z}  +  \vec{\delta},
\end{equation}
the covariance matrix elements according to  the two sides of the detailed-balance condition Eq. (\ref{det2}) follow explicitly as 
\begin{equation} \label{det4a}
    \langle \delta_o \delta_p \rangle^{\rm ls} = E_{op}(t) + 
   (\delta_{ok} -   {\rm e}^{ - t A }_{ok})   (\delta_{pl} -   {\rm e}^{ - t A }_{pl}) E^{\circ}_{kl} 
    \end{equation}
and
\begin{eqnarray} \label{det4b}
&    \langle \delta_o \delta_p \rangle^{\rm rs} = W_{oi} E_{ij}(t) W_{jp}    \\
   &+ (\delta_{ok} -  W_{ol} {\rm e}^{ - t A }_{lm} W_{mk})   (\delta_{pn} - W_{pq}  {\rm e}^{ - t A }_{qr} W_{rn}) E^{\circ}_{kn}.  \nonumber
    \end{eqnarray}
Since a Gaussian distribution is completely specified by its covariance matrix, detailed balance is satisfied if
$  \langle \delta_o \delta_p \rangle^{\rm ls} =   \langle \delta_o \delta_p \rangle^{\rm rs} $ holds.
Similar to previous treatments  \cite{Kampen1957a,Kampen1957b,Graham1971}, we expand the detailed-balance condition in  time. 
From Eq. 	(\ref{sol3}) we obtain
\begin{equation} \label{det5}
   E_{ij}(t) = 2 t C_{ij}  - t^2 ( A _{ik}   C_{kj} +  A _{jk}   C_{ki})  + {\cal O}(t^3).
       \end{equation}
To second order in $t$ we  thus obtain 
\begin{equation} \label{det6a}
    \langle \delta_o \delta_p \rangle^{\rm ls} = 2 t C_{op} +
  t^2 (A _{ok} E^{\circ}_{kl}  A_{pl} - A _{ok}   C_{kp} -  A _{pk}   C_{ko})
     \end{equation}
and
\begin{eqnarray} \label{det6b}
 &    \langle \delta_o \delta_p \rangle^{\rm rs} = 2 t W_{oi} C_{ij} W_{jp} \\ \nonumber
&+  t^2 W_{ol}  (A _{lm} W_{mn}  E^{\circ}_{nt} W_{ts}   A_{rs} - A _{lk}   C_{kr} -  A _{rk}   C_{kl}) W_{rp}.  \\ \nonumber
     \end{eqnarray}
From enforcing the detailed-balance condition term by term  in powers of $t$,  we obtain the two conditions
\begin{equation} \label{det7a}
   C_{op} = W_{oi} C_{ij} W_{jp}
     \end{equation}
and
\begin{eqnarray} \label{det7b}
 &  A _{ok} E^{\circ}_{kl}  A_{pl} - A _{ok}   C_{kp} -  A _{pk}   C_{ko} \\ \nonumber
& =  W_{ol}  (A _{lm} W_{mn}  E^{\circ}_{nt} W_{ts}   A_{rs} - A _{lk}   C_{kr} -  A _{rk}   C_{kl}) W_{rp}, \\  \nonumber
     \end{eqnarray}
which are consistent with previous results  \cite{Kampen1957a,Kampen1957b,Graham1971}.
To evaluate the conditions Eqs. (\ref{det7a}) and  (\ref{det7b}), it is useful  to express  all matrices as products of   particle matrices and 
$2 \times 2$ submatrices as introduced in Sect. \ref{EntropyProd}. With this, the matrix $W$  in Eq. (\ref{W}) can be written as 
\begin{equation} 	\label{det8a}
W_{\alpha\beta} = \delta_{\alpha\beta}(s-r),
     \end{equation}
where $s$ and $r$ are  $2 \times 2$ matrices  defined in Eq.  (\ref{2by2}) and where greek indices number particles.  
To test the consequences of condition Eq. (\ref{det7a}), we expand the random correlation matrix in the complete set of 
$2 \times 2$ submatrices according to 
\begin{equation}
	\label{det8b}
C_{\alpha \beta} = s C_{\alpha \beta}^{\rm s}   + r C_{\alpha \beta}^{\rm r}  r+ 
ut C_{\alpha \beta}^{\rm ut}  + us C_{\alpha \beta}^{\rm us}.
\end{equation}
Inserting Eqs. (\ref{det8a}) and  (\ref{det8b})  into Eq. (\ref{det7a}), we  obtain by using the matrix product properties Eq. (\ref{2by2c})
that $C_{\alpha \beta}^{\rm ut}= 0 = C_{\alpha \beta}^{\rm us}$, i.e. the random correlation matrix cannot contain components that couple
momenta and positions. We will later see that this condition is indeed satisfied for a large class of Hamiltonian models.

To evaluate the consequences of 
the condition Eq. (\ref{det7b}), we  need to specify the symmetry of the Hamiltonian model. 
As in Eq. (\ref{Ham2}),  the  Hamiltonian matrix shall describe a  Newtonian system where positions and momenta  are decoupled
and furthermore momentum contributions are diagonal in the particles indices,
\begin{equation}
	\label{det9}
	H_{\alpha' \epsilon}   =   s h_{\alpha' \epsilon}  +  r \delta_{\alpha' \epsilon} /m_{\alpha'}.
	\end{equation}
The friction matrix is copied from Eq.(\ref{Gamma})  and only acts on momentum degrees of freedom, 
\begin{equation} \label{det10}
 \Gamma_{\alpha \epsilon'} = r \gamma_{\alpha \epsilon'}   / m_{\epsilon'},
\end{equation}
 in the following we  explicitly allow the matrix $\gamma_{\alpha \epsilon'}$ also to be asymmetric. 
Using the inverse Hamiltonian from Eq. (\ref{Ham3}) and Eq. (\ref{det10}),  the
Lyapunov-Boltzmann  condition  Eq. (\ref{LyapBoltz}) yields  the equilibrium random correlation matrix 
\begin{equation}
\label{det11}
    2 C_{\alpha \epsilon }/(k_BT)=( \gamma_{\alpha \epsilon} +  
     \gamma_{\epsilon \alpha })r,
\end{equation}
which only acts on momentum degrees of freedom. We conclude that the first detailed-balance condition  Eq. (\ref{det7a})
is automatically satisfied for a Newtonian system that is described by the Boltzmann distribution and thus satisfies the Lyapunov-Boltzmann condition 
Eq. (\ref{LyapBoltz}). 

We next probe condition Eq. (\ref{det7b}). According to the  definition used in the main text, in equilibrium  $E^{\circ}_{kl}  = E^{\bullet}_{kl}  = k_BT H^{-1}_{kl}$ 
holds and thus the stationary covariance matrix $E^{\circ}_{kl}$
does not couple momentum and position coordinates, therefore  condition Eq. (\ref{det7b}) simplifies to 
\begin{eqnarray} \label{det12}
 &  A _{ok} E^{\bullet}_{kl}  A^T_{lp} - A _{ok}   C_{kp} -     C^T_{ok}  A^T _{kp} \\ \nonumber
& =  W_{ol}  (A _{lm}  E^{\bullet}_{ms}   A^T_{sr} - A _{lk}   C_{kr} -  C^T_{lk} A^T _{kr}   ) W_{rp}, \\  \nonumber
     \end{eqnarray}
where the superindex $T$ denotes the transpose of the matrix including the $2 \times 2$ submatrices.
Equation (\ref{det12})  holds if the left side is a matrix that does not couple momenta and positions. 
From  the expression for $A$,  Eq. (\ref{A2}),  and $ U_{\alpha \gamma} =    u \delta_{\alpha \gamma}$ we obtain
\begin{equation}
	\label{det13}
	A_{\alpha' \epsilon'}   =   - us h_{\alpha' \epsilon'}  -   ur \delta_{\alpha' \epsilon'} /m_{\alpha'} + 
	 r \gamma_{\alpha' \epsilon'}   / m_{\epsilon'}.
	\end{equation}
Replacing  $E^{\bullet}_{kl} $ by 
the explicit result for $H^{-1}_{kl}$ from Eq. (\ref{Ham3}), and using  the results for $C$ and $A$ from Eqs. 
(\ref{det11}) and (\ref{det13}) we finally obtain for the left hand side of Eq. (\ref{det12})
\begin{eqnarray} \label{det14}
 &  r h_{\alpha' \beta'} + s  \delta_{\alpha' \beta'} m_{\beta'}^{-1} \\ \nonumber
& + r \left[ ( \gamma_{ \epsilon \alpha'} -    \gamma_{\alpha' \epsilon  })  \gamma_{\beta' \epsilon  }
 +  ( \gamma_{ \epsilon \beta'} -    \gamma_{\beta' \epsilon  })  \gamma_{\alpha' \epsilon  } \right] m_{\epsilon}^{-1} /2
\\ \nonumber
& + ru  ( \gamma_{ \alpha' \beta'} -    \gamma_{\beta' \alpha'  }) m_{\beta'}^{-1} /2
+   ur  ( \gamma_{ \alpha' \beta'} -    \gamma_{\beta' \alpha'  }) m_{\alpha'}^{-1} /2.
\\  \nonumber
     \end{eqnarray}
For an asymmetric friction matrix $\gamma_{ \alpha \beta}$, this expression contains terms that couple positions and momenta, 
 these are the two terms proportional to $ru$ and $ur$, and therefore the second detailed-balance condition Eq. (\ref{det7b})  is not satisfied, 
 even if the system obeys the Boltzmann distribution. If the friction matrix $\gamma_{ \alpha \beta}$
 is  symmetric, the terms proportional to $ru$ and $ur$
 disappear and the detailed-balance condition is satisfied.
 We conclude that for a symmetric friction matrix the condition of detailed balance and the definition of equilibrium we use in the main text, based on the
 Boltzmann distribution, are equivalent. Since none of the effects we study in this paper depend on the presence of asymmetries in the friction matrix, 
 it is permissible and convenient  to use the Boltzmann definition for equilibrium, which we primarily do since it is much easier to implement. 
 The presence of an asymmetric friction matrix means that the principle of equal actio and reactio is broken on the level of the Langevin equation, 
 this therefore constitutes a distinct  NEQ scenario.

\section{Condition of vanishing probability current  for overdamped particle dynamics}  \label{AppCnew}

In Sect. \ref{detbal} we show that the probability current is generally nonzero for underdamped particles. Here we consider
overdamped particle motion and show that the probability  current vanishes only for a symmetric friction matrix.
To procced,
we rewrite  the equation of motion  expressed  in terms of particle coordinates $ y_\alpha(t) = (x_\alpha(t), p_\alpha(t))^T$,
  Eq. (\ref{Langevin2}), as a second-order  differential equation as
\begin{equation}
	\label{Langmassive1}
m_{\alpha'}   \, \ddot{x}_{\alpha'}(t) =  - \gamma_{\alpha' \beta}  \dot{x}_\beta(t) - h_{\alpha' \beta}  x_\beta (t)
+ \phi_{\alpha'  \beta}  F_\beta (t),
\end{equation}
where the Hamiltonian matrix $h_{\alpha' \beta}$ only acts on positions and  the
friction matrix $ \gamma_{\alpha' \beta}$ only acts on momenta.
$ \phi_{\alpha'  \beta}$ is the $N\times N$ random coupling strength matrix. 
Setting all  masses $m_\alpha$  to zero,  we obtain
\begin{equation}
	%\label{Lang}
  \dot{x_\beta}(t)=  -  \gamma^{-1}_{\alpha \beta} h_{ \beta \gamma }  x_\gamma (t)
+  \gamma^{-1}_{\alpha \beta}   \phi_{ \beta \gamma }  F_\gamma (t),
\end{equation}
which we can rewrite in a form similar to Eq. (\ref{Langevin})  as 
\begin{equation}
	% \label{Langevin}
\dot{x}_\alpha(t)= -A^{\rm od}_{\alpha \beta} x_\beta (t) + \Phi^{\rm od}_{\alpha \beta} F_\beta  (t),
\end{equation}
where the overdamped versions of $A$ and $\Phi$ are $N\times N$ matrices given by 
$A^{\rm od}_{\alpha \gamma}   =   \gamma^{-1}_{\alpha \beta} h_{ \beta \gamma } $
and $ \Phi^{\rm od}_{\alpha \gamma} =  \gamma^{-1}_{\alpha \beta}   \phi_{ \beta \gamma }$.
The overdamped version of Eq. (\ref{CCC}) is 
$C^{\rm od}_{kl}=\Phi^{\rm od}_{km}\Phi^{\rm od}_{lm}$.
The overdamped version of the Lyapunov-Boltzmann  equation Eq. (\ref{LyapBoltz}) turns out to be
\begin{equation}
2 C^{\rm od}_{\alpha  \beta }/(k_BT) = \gamma^{-1}_{\alpha \beta} +  \gamma^{-1}_{ \beta \alpha}
\end{equation}
and constitutes the relation between the  random strength matrix  and  the friction matrix for overdamped systems.
The overdamped version of the condition of vanishing probability current follows from   Eq. (\ref{balance4})  as 
\begin{equation}
C^{\rm od}_{\alpha  \beta }/(k_BT) = \gamma^{-1}_{\alpha \beta} 
\end{equation}
and, since $ C^{\rm od}_{\alpha  \beta }$ is symmetric by construction,
can  only  be satisfied if the friction matrix $\gamma_{\alpha \beta} $
is symmetric. Thus, if  $\gamma_{\alpha \beta} $ is not symmetric,
the probability current is non-zero  even if the distribution is given by the Boltzmann distribution. We conclude that the condition of 
vanishing probability current and the equilibrium definition we use in this paper, namely that the
stationary distribution is given by the Boltzmann distribution, see Eqs. (\ref{Boltz}) and (\ref{Ebullet}),
 are not equivalent even in the overdamped limit. In fact,  
 the condition of vanishing probability current  is more restrictive since it precludes asymmetric friction matrices,
 which is not necessary to obtain Boltzmann distributions. Thus, there is a class of overdamped particle models that exhibit
 stationary Boltzmann distributions yet that exhibit non-zero probability currents in the stationary state.

\section{Fluctuation-dissipation theorem} 
 \label{AppDnew}
Here  we  demonstrate for a specific  two-particle model with an asymmetric friction matrix,
that the fluctuation-dissipation theorem can be  satisfied even when the system, as demonstrated
in Appendix \ref{AppEnew},  does not obey detailed balance.
To proceed, we consider a system of two massive particles that are coupled by a general friction matrix. 
Following the notation of Eq. (\ref{Langmassive1}) we write
\begin{widetext}
\begin{equation}
	\label{Langmassive1app}
m_1   \, \ddot{x}_1(t) =  - \gamma_{1}  \dot{x}_1(t) - \gamma_{12}  \dot{x}_2(t) + \phi_{1}  F_1 (t) + \phi_{12}  F_2 (t),
%+ f[x_1(t)],
\end{equation}
\begin{equation}
	\label{Langmassive2app}
m_2   \, \ddot{x}_2(t) =  - \gamma_{2}  \dot{x}_2(t) - \gamma_{21}  \dot{x}_1(t) + \phi_{2}  F_2 (t) + \phi_{21}  F_1 (t).
\end{equation}
%where $f[x_1(t)]$ is a general non-linear function.
Note that for ease of calculation we omit any positional couplings between the particles.
 Similar to the active-particle  model in Sect.  \ref{activeparticles},
where the two particle positions  are coupled,
the Langevin equation for the second particle can be    solved and gives 
\begin{equation}
\dot{x}_2(t) = \int_{-\infty}^t {\rm d}t' {\rm e}^{-(t-t')\gamma_2/m_2}
\left[ - \frac{\gamma_{21}}{m_2} \dot{x}_1(t') +  \frac{\phi_2}{m_2} F_2(t') +  \frac{\phi_{21}}{m_2}   F_1(t') \right].
\end{equation}
Inserting this solution into the Langevin equation  for the first particle,
we obtain  the generalized Langevin equation 
\begin{equation} \label{gleapp}
m_1 \ddot{x_1} (t) = -
\int_{-\infty}^\infty {\rm d}t' \Gamma(t-t') \dot{x}_1(t') +F_R(t). 
\end{equation}
The memory function that appears in  Eq. (\ref{gleapp})  is given by 
\begin{equation} \label{memoryapp}
\Gamma (t) = \theta(t)  \left[ 2 \gamma_1 \delta(t)  - \frac{\gamma_{12}  \gamma_{21}}{m_2}
   {\rm e}^{- t \gamma_2/m_2}\right],
\end{equation}
where $\theta(t)$ denotes the Theta function with the properties
$\theta(t)=1$ for $t>0$ and $\theta(t)=0$ for $t<0$, which makes 
the memory function  single-sided. 
The noise $F_R(t)$   in  Eq. (\ref{gleapp})  is given by 
\begin{equation}
F_R (t) =  \phi_1  F_1(t) +  \phi_{12}  F_2(t) -  \int_{-\infty}^t {\rm d}t' 
{\rm e}^{-(t-t')\gamma_2/m_2}
\left[   \frac{  \gamma_{12} \phi_2}{m_2} F_2(t') +  \frac{ \gamma_{12} \phi_{21}}{m_2}   F_1(t') \right].
\end{equation}
The auto-correlation function of the random noise  
 $C_{\rm FF}(t)=\langle F_R(0) F_R(t) \rangle$ follows as
\begin{equation} \label{CFFapp}
 C_{\rm FF}(t)   =   
 2 (\phi_1^2 + \phi_{12}^2)  \delta(t) -[2 \gamma_{12}(   \phi_2\phi_{12} + \phi_1\phi_{21})  -\gamma_{12}^2(\phi_2^2+ \phi_{21}^2)/\gamma_2]
     {\rm e}^{-|t|  \gamma_2/m_2}.
\end{equation}
Defining the Langevin equation in analogy to  Eq. (\ref{Langevin2}) in terms of the particle momenta 
$p_{\alpha'}(t)=m_{\alpha'}  v_{\alpha'}(t)$ as 
$\dot{p}_\alpha (t)  = -A_{\alpha \beta} p_\beta(t) +\phi_{\alpha \beta }F_\beta  (t)$
where  $A_{\alpha \beta'}= \gamma_{\alpha \beta'}/m_{\beta'}$ and  $H_{\alpha \beta'}= \delta_{\alpha \beta'}/m_{\beta'}$,
we obtain from Eq. (\ref{LyapBoltz0}) the Boltzmann-Lyapunov equation  
\begin{equation}
2 C_{\alpha  \beta } =2 \phi_{km}\phi_{lm}  =  k_BT ( \gamma_{\alpha \beta} + \gamma_{ \beta \alpha} ),
\end{equation}
or, explicitly, 
 \begin{equation} \label{Uapp}
 2 \begin{pmatrix}
 \phi_1^2+ \phi_{12}^2    &   \phi_1\phi_{21} + \phi_2\phi_{12}   \\
 \phi_1\phi_{21} + \phi_2\phi_{12}  &  \phi_2^2+ \phi_{21}^2   
\end{pmatrix}	
=k_BT 
\begin{pmatrix}
2 \gamma_1   & \gamma_{12} + \gamma_{21}   \\
 \gamma_{12} + \gamma_{21}   &  2  \gamma_2 
\end{pmatrix}.
\end{equation}
\end{widetext}
Provided  Eq. (\ref{Uapp}) holds,  it is easy to see that 
Eqs. (\ref{memoryapp}) and Eq. (\ref{CFFapp})
satisfy the fluctuation-dissipation theorem \cite{Risken}
 \begin{equation}
  C_{\rm FF}(t)  =k_BT  \Gamma (|t|),
  \end{equation}
 even if the friction matrix $ \gamma_{\alpha \beta}$  is asymmetric. 
In contrast and as shown in Appendix \ref{AppEnew},  
 the more restrictive detailed-balance condition is only satisfied 
  if the friction matrix $ \gamma_{\alpha \beta}$ is symmetric.
This suggests  that the definition of equilibrium we use in this paper, 
namely that the stationary distribution corresponds to the Boltzmann distribution, coincides with the 
fluctuation-dissipation theorem even for asymmetric   $ \gamma_{\alpha \beta}$. On the other hand, 
the detailed-balance condition is more restrictive and can only be satisfied if the friction matrix is symmetric.

\section{Equivalenz of free entropy and Kullback-Leibler entropy}  \label{AppGnew}

The multidimensional expression for the Kullback-Leibler entropy reads \cite{Kullback1951,Schloegl1971}
\begin{equation}
\label{KL1}
{\cal S}_{\rm KL}    =   -  \int {\rm d} \vec{z}  \rho(\vec{z})  \ln \left[ 
\frac{ \rho^\circ (\vec{z})}{  \rho(\vec{z})} \right],
\end{equation}
where the Gaussian probability distribution $\rho( \vec{z})$  and the 
Gaussian stationary probability distribution $\rho^\circ( \vec{z}) $ are
given by 
\begin{equation}
\rho( \vec{z})  =  {\cal N}^{-1} \exp\left(-    z_i  E^{-1}_{ij} z_j /2 \right),
 \end{equation}
\begin{equation}
\rho^\circ( \vec{z})  =  {\cal N^\circ}^{-1} \exp\left(-    z_i  E^{\circ -1}_{ij} z_j /2  \right)
 \end{equation}
with the normalization constants %
$ {\cal N}  = \sqrt{(2\pi)^{2N} \det E}$ and $ {\cal N}^\circ  = \sqrt{(2\pi)^{2N} \det E^\circ}$.
Note that we dropped all time dependencies for simplicity. Using the inequality
\begin{equation}
\label{KL2}
-  \ln \left[ 
\frac{ \rho^\circ (\vec{z})}{  \rho(\vec{z})} \right] \geq 1- \frac{ \rho^\circ (\vec{z})}{  \rho(\vec{z})}
\end{equation}
and the fact that $\rho( \vec{z})$  and $\rho^\circ( \vec{z}) $ are normalized, it immediately follows that 
\begin{equation}
\label{KL3}
{\cal S}_{\rm KL}      \geq 0 
\end{equation}
and that ${\cal S}_{\rm KL}    =  0$ for   $\rho( \vec{z}) = \rho^\circ( \vec{z})$. 
A short calculation shows that also $\delta {\cal S}_{\rm KL} / \delta \rho( \vec{z})   =  0$ for   
$\rho( \vec{z}) = \rho^\circ( \vec{z})$, 
thus, the Kullback-Leibler is minimal  for $\rho( \vec{z}) = \rho^\circ( \vec{z})$. Performing the Gaussian 
integrals in Eq. (\ref{KL1}), one immediately obtains
\begin{eqnarray}
\label{KL4}
&{\cal S}_{\rm KL}    =  E^{\circ -1}_{ij} E_{ij}/2 - N 
-  \ln \left[ (2\pi)^{2N} \det E \right] /2\nonumber \\
& +  \ln\left[ (2\pi)^{2N} \det E^\circ \right]  /2.
\end{eqnarray}
Comparison of  the  Kullback-Leibler entropy
Eq. (\ref{KL4}) with the non-equilibrium free entropy expression Eq. (\ref{Sfree3})
shows that the two expressions are identical except the sign and terms
that do not depend on  covariance matrix elements $E_{ij}$.

%\section{Derivation of time-dependent entropy} \label{AppD}
\section{Non-stationary  entropy production  of heat  reservoirs} \label{AppE}

We start from   the expression for the time-dependent
 entropy production of all reservoirs Eq. (\ref{Sprod5}), which 
 is based on the result  for the heat fluxes  between the reservoirs 
 and the system in Eq. (\ref{Sprod2}). We recall  that the calculation
 of the reservoir heat fluxes is based on 
 the general Newtonian Hamiltonian as defined by Eq. (\ref{Ham2})
with diagonal momentum friction as defined by Eq. (\ref{Gamma})
 and the NEQ  model defined by  the  diagonal
 noise correlation matrix Eq. (\ref{LyapBoltz4}).

Using the result for the time derivative of 
the covariance matrix Eq. (\ref{covT})
together with the explicit expression for the $A$ matrix Eq. (\ref{A2})
and the expansion of the covariance matrix $E$ Eq. (\ref{Sprod3}),
we obtain for the time derivative of the squared particle momenta
\begin{equation}
	\label{psquare}
 \frac{ {\rm d}    \langle p^2_{\alpha'} (t) \rangle }{ {\rm d} t}
 = 2C_{\alpha' \alpha'} - \frac{2\gamma_{\alpha'}}{m_{\alpha'}}\langle p^2_{\alpha'} (t) \rangle
 - 2  h_{\alpha' \epsilon}  \langle  p_{\alpha'} x_\epsilon  \rangle.
 \end{equation}
Inserting this into the total reservoir entropy production   Eq. (\ref{Sprod5})
we obtain
\begin{equation}
	\label{Sprod7b}
\frac{ \dot{\cal S}_{\rm res}(t) }{k_B} 
 =   \frac{ \beta_\alpha \gamma_\alpha  }{m_{\alpha}} \left(    
 \frac{  \langle p^2_{\alpha} (t) \rangle }{ m_{\alpha}  } 
 -  \frac{1}{ \beta_\alpha  } \right),
  \end{equation}
where the non-primed index $\alpha$ is summed over.
Clearly, in an equilibrium stationary state, where the kinetic energy of each particle obeys
$ \langle p^2_{\alpha} (t) \rangle / ( m_{\alpha}) = 1/\beta_\alpha$ and 
$\beta_\alpha=\beta$,
the entropy production vanishes. In a stationary state we can replace 
$ \langle p^2_{\alpha} (t) \rangle$ by  the expression
Eq. (\ref{Lyapt2}) and thereby recover Eq. (\ref{Sprod6}), so our calculation thus far  is consistent.
We will now test whether the  stationary NEQ  distribution
constitutes an extremum of the total entropy production, which according to Eq. (\ref{STOT2}) is given by
\begin{equation}
	\label{Sprod8b}
\dot{\cal S}_{\rm tot}(t) = \dot{\cal S}(t)+ \dot{\cal S}_{\rm res}(t),
  \end{equation}
where $ \dot{\cal S}(t)$ denotes the Shannon entropy production of the 
system distribution Eq. (\ref{entropy3}) and $\dot{\cal S}_{\rm res}(t)$ is given by
 Eq. (\ref{Sprod7b}). As an explicit calculation shows, 
the derivative of the reservoir entropy production Eq. (\ref{Sprod7b}) with respect to the 
covariance matrix  gives
\begin{equation}
	\label{Sprod9b}
\frac{\partial\dot{\cal S}_{\rm res}(t)}{\partial E_{ij}}=
\Gamma_{ik}  C^{-1}_{km}   \Gamma_{mj},
  \end{equation}
which using  Eqs.  (\ref{Gamma}) and (\ref{LyapBoltz4})
 is a diagonal matrix with momentum entries 
$\beta_{\alpha'} \gamma_{\alpha'} /m^2_{\alpha'}$.
For the system entropy production we obtain from Eq. (\ref{entropy3})  
\begin{equation}
	\label{Sprod10b}
\frac{\partial \dot{\cal S}(t)}{\partial E_{ij}}=
- E^{-1}_{ik}(t)   C_{km}   E^{-1}_{mj}(t)
  \end{equation}
and thus for the total entropy 
\begin{equation}
	\label{Sprod11b}
\frac{\partial \dot{\cal S}_{\rm tot}(t)}{\partial E_{ij}}=
\Gamma_{ik}  C^{-1}_{km}   \Gamma_{mj}- E^{-1}_{ik}(t)   C_{km}   E^{-1}_{mj}(t).
  \end{equation}
Multiplying by $C_{li}$ we obtain
\begin{equation}
	\label{Sprod12b}
C_{li} \frac{\partial \dot{\cal S}_{\rm tot}(t)}{\partial E_{ij}}=
\Gamma_{lk}     \Gamma_{kj}   - C_{li}  E^{-1}_{ik}(t)   C^{-1}_{km}   E^{-1}_{mj}(t),
  \end{equation}
which vanishes when 
\begin{equation}
	\label{Sprod13b}
\Gamma_{lk}     \Gamma_{kj}   = C_{li}  E^{-1}_{ik}(t)   C^{-1}_{km}   E^{-1}_{mj}(t)
  \end{equation}
holds. Taking the square root and multiplying by $E$
we obtain 
\begin{equation}
	\label{Sprod14b}
\Gamma_{lk}  E_{ki}(t)     = C_{li}    
  \end{equation}
as the equation which determines the extremum of the total entropy production.
 It turns out that  that Eq. (\ref{Sprod14b}) is satisfied by the equilibrium
 stationary distribution $E^\bullet_{ki}$, therefore for an equilibrium system the total entropy 
 production as defined in Eq. (\ref{Sprod8b}) is extremal in the equilibrium state. 

It is easy to verify that for the stationary NEQ  distribution
$E^\circ_{ki}$ with non zero matrix elements $\langle p_i p_j \rangle^\circ$
and $\langle x_i p_j \rangle^\circ$ for $i  \neq j$, 
and $\Gamma$ and $C$ given  by Eqs. (\ref{Gamma})
 and  (\ref{LyapBoltz4}), Eq. (\ref{Sprod14b})  is not satisfied. 
This implies that the total entropy production Eq. (\ref{Sprod8b}),
which is the sum of the 
non-stationary reservoir entropy production 
 Eq. (\ref{Sprod5}) or  (\ref{Sprod7b}) 
 (calculated from the heat fluxes between reservoirs and the system)
 and the system distribution entropy 
 production  Eq. (\ref{entropy3}),  does not yield the exactly calculated stationary 
 distribution $E^\circ_{ki}$ at its extremum
 and thus differs from the total entropy production 
 constructed from the sum of the  free entropy and the stationary reservoir entropy production
 according to Eq. (\ref{STOT2}).
 We conclude  that non-stationary
 entropy contributions that  presumably  stem from internal  reservoir degrees of freedom 
 render  the expressions Eqs. (\ref{Sprod5}) and   (\ref{Sprod7b}) incomplete.
 These internal reservoir degrees could be included  by using 
 microscopic models for the heat reservoirs.

\end{appendix}

%\bibliography{noneqinter.bib}

%merlin.mbs apsrev4-1.bst 2010-07-25 4.21a (PWD, AO, DPC) hacked
%Control: key (0)
%Control: author (72) initials jnrlst
%Control: editor formatted (1) identically to author
%Control: production of article title (-1) disabled
%Control: page (0) single
%Control: year (1) truncated
%Control: production of eprint (0) enabled
%

\end{document}